\pdfoutput=1
\RequirePackage{ifpdf}
\ifpdf 
\documentclass[pdftex]{sigma}
\else
\documentclass{sigma}
\fi

\newcommand{\sgn}{\mathop{\rm sgn}}

\newtheorem{thm}{Theorem}[section]
\newtheorem{prop}[thm]{Proposition}
\newtheorem{Lemma}[thm]{Lemma}

{\theoremstyle{definition}

}

\numberwithin{equation}{section}

\def\F{{\mathcal{F}}}

\def\CC{{\mathbb{C}}}
\newcommand{\ZZ}{{\mathbb Z}}
\def\TT{{\mathbb{T}}}
\def\FF{{\rm F}}

\def\EE{{\rm E}}
\def\FF{{\rm F}}
\def\hEE{\hat{\rm E}}
\def\hFF{\hat{\rm F}}
\def\hk{\hat{k}}
\def\TT{{T}}
\def\tEE{\tilde{\rm E}}
\def\tFF{\tilde{\rm F}}
\def\tTT{\tilde{T}}
\def\hTT{\hat{T}}

\def\Ee{{\sf e}}
\def\E{\mathcal{E}}
\def\hF{\hat{F}}
\def\hE{\hat{E}}

\def\sk#1{\left(#1\right)}

\def\Pep{{P}^+_e}
\def\Pem{{P}^-_e}

\def\Pfp{{P}^+_f}
\def\Pfm{{P}^-_f}

\def\DYAn{\mathcal{D}Y(\mathfrak{gl}_{N})}
\def\DYBn{\mathcal{D}Y(\mathfrak{o}_{2n+1})}
\def\DYCn{\mathcal{D}Y(\mathfrak{sp}_{2n})}
\def\DYDn{\mathcal{D}Y(\mathfrak{o}_{2n})}
\def\DYg{\mathcal{D}Y(\mathfrak{g})}
\def\Bc{\mathcal{B}}

\def\Ac{\mathcal{A}}
\def\oAc{\overline{\mathcal{A}}}
\def\Cc{\mathcal{C}}

\def\Dc{\mathcal{D}}
\def\oDc{\overline{\mathcal{D}}}
\def\Xc{\mathcal{X}}
\def\oXc{\overline{\mathcal{X}}}

\def\DYGLN{\mathcal{D}Y(\mathfrak{gl}_{N})}

\def\DAal{\mathcal{DA}}
\def\Bal{\mathcal{B}}
\def\DBal{\mathcal{DB}}
\def\Cal{\mathcal{C}}
\def\DCal{\mathcal{DC}}

\def\DDal{\mathcal{DD}}

\def\fgo{\mathfrak{f}}

\def\ot{\otimes}
\def\cF{{\sf F}}
\def\cE{{\sf E}}

\def\IA{I_{\mathcal A}}
\def\IB{I_{\mathcal B}}
\def\IC{I_{\mathcal C}}
\def\ID{I_{\mathcal D}}
\def\Ig{I_{\mathfrak{g}}}

\begin{document}
\allowdisplaybreaks

\newcommand{\arXivNumber}{2006.01579}

\renewcommand{\PaperNumber}{120}

\FirstPageHeading

\ShortArticleName{Gauss Coordinates vs Currents for the Yangian Doubles of the Classical Types}

\ArticleName{Gauss Coordinates vs Currents\\ for the Yangian Doubles of the Classical Types}

\Author{Andrii~LIASHYK~$^\dag$ and Stanislav Z.~PAKULIAK $^\ddag$}

\AuthorNameForHeading{A.~Liashyk and S.Z.~Pakuliak }

\Address{$^\dag$~Skolkovo Institute of Science and Technology, Moscow, Russia}
\EmailD{\href{mailto:a.liashyk@gmail.com}{a.liashyk@gmail.com}}

\Address{$^\ddag$~Steklov Mathematical Institute of Russian Academy of Sciences,\\
\hphantom{$^\ddag$}~8 Gubkina Str., Moscow, 119991, Russia}
\EmailD{\href{mailto:stanislav.pakuliak@jinr.ru}{stanislav.pakuliak@jinr.ru}}

\ArticleDates{Received June 03, 2020, in final form November 15, 2020; Published online November 22, 2020}

\Abstract{We consider relations between Gauss coordinates of $T$-operators for the Yangian doubles of the classical types corresponding to the algebras $\mathfrak{g}$ of $A$, $B$, $C$ and $D$ series and the current generators of these algebras. These relations are important for the applications in the quantum integrable models related to $\mathfrak{g}$-invariant $R$-matrices and construction of the Bethe vectors in these models.}


\Classification{82B23; 81R12; 81R50; 17B80}

\section{Introduction}

Nested algebraic Bethe ansatz is a powerful tool for investigation of the quantum
integrable models associated with $\mathfrak{g}$-invariant $R$-matrices. Since pioneering papers
\cite{KulRes81,KulRes83} this method was mostly developed for the case when
$\mathfrak{g}$ belongs to the type $A$ algebras and their supersymmetric generalizations.

To deal with nested algebraic Bethe ansatz one can identify the monodromy matrix elements
of some quantum integrable model related to $\mathfrak{g}$- or $U_q(\mathfrak{g})$-invariant
$R$-matrices with the generating
series of the elements of the Yangian doubles or quantum affine algebras \cite{Dr88}.
These generating series gathered into $T$-operators satisfy the same commutation relations defined by some
$\mathfrak{g}$- or~$U_q(\mathfrak{g})$-invariant $R$-matrix as monodromy matrix elements
of the corresponding quantum integrable model do.
These realizations of the Yangian doubles or quantum affine algebras are known as $RTT$-realization
\cite{RS1990}. The same algebras have so called ``current'' or ``new'' rea\-li\-zation~\cite{D88}.

For the type $A$ algebras an explicit relations between generating series in the $RTT$-formu\-la\-tion and
the currents in the ``new'' realization were found in \cite{DF93}.
Recent results published in the papers \cite{JLM18,JLM19,JLM19a,JLY18} describes similar
equivalences for the other type algebras.

The main object we are considering in this paper is a Yangian double $\mathcal{D}Y(\mathfrak{g})$
described in~\cite{JLY18} as the algebra which is topologically generated by the modes of matrix entries
of the $T$-operators and the central element.
 In the present paper we will not need a central extension of the Yangian double and will not
 discuss the topological aspects of this algebra.
 Readers interested in this subject may be addressed to the paper~\cite{JLY18}
and references therein. But the main result of~the~latter paper was the proof of isomorphism
between $RTT$ and ``current'' realizations of the Yangian doubles using Gauss decomposition of
$T$-operators. This proof yields, in particular, the~explicit relations between Gauss coordinates
and currents corresponding to the simple roots of~the~algebra~$\mathfrak{g}$.

A significant development of nested algebraic Bethe ansatz
was achieved in the papers \cite{EKhP07,HutLPRS17a,KhP-Kyoto}
when the main objects of this method, so called Bethe vectors, were constructed in terms of
the current generators of the Yangian doubles and quantum affine algebras. This construction
uses the projections onto intersections of the different types Borel subalgebras in these infinite
dimensional algebras corresponding to their $RTT$ and ``current'' realizations.
To use method of projections in nested algebraic Bethe ansatz one has to express all Gauss coordinates through
current generators. This can be realized through the same projection method and
allowed in~\cite{KhP-Kyoto} to obtain the explicit formulas for the off-shell Bethe vectors in terms
of the monodromy matrix elements for quantum integrable models associated
with $U_q(\mathfrak{gl}_N)$-invariant $R$-matrices.
Analogous results were obtained
in \cite{HutLPRS17a} for the models related to the supersymmetric $\mathfrak{gl}(m|n)$-invariant
$R$-matrices.

The goal of the present paper is to extend the results of \cite{JLY18} and to establish the relations between all Gauss coordinates of $T$-operators and the currents of the Yangian doubles for all types classical algebras. This correspondence is described in Proposition~\ref{cuvsgc}.

The paper is composed as follows. Section~\ref{RTT} describes the Yangian double
$\mathcal{D}Y(\mathfrak{g})$ in the $RTT$ formulation. In Section~\ref{GaussCoor}
we introduce Gauss coordinates of the $T$-operators and present current realizations
of the Yangian double $\mathcal{D}Y(\mathfrak{g})$ following the paper \cite{JLY18}.
Since we used a different Gauss decomposition, we cannot use directly the commutation relations
between currents given in~\cite{JLY18}. The main new results of this section are the relations between
Gauss coordinates~\eqref{oth-ge}, \eqref{sp-ge} and \eqref{o2n-ge} of $T$-operators for the
Yangian doubles $\DYBn$, $\DYCn$ and $\DYDn$ respectively. These formulas
are direct consequence of the equality
\eqref{ident} and the results obtained in \cite{LPRS19}. For the sake of completeness
we recall these results in the Section~\ref{3.1}. Following
the ideas developed in the paper~\cite{EKhP07}
we introduce projections onto intersections of the different types Borel subalgebras for
$RTT$ and ``current'' realizations of the Yangian doubles in~Section~\ref{GCandproj}.
Section~\ref{GCvscur} is devoted to the formulation of the main result of this paper
describing the explicit relations between all Gauss coordinates of $T$-operators and
the currents. Appendix~\ref{ApA} contains the proof of Proposition~\ref{cuvsgc} and
Appendix~\ref{ApB} describes properties of the automorphism of the Yangian double
$\DYAn$ given by the formulas \eqref{hFF}--\eqref{hEE}.

\section[Yangian double DY(g) in the RTT formulation]
{Yangian double $\boldsymbol{\mathcal{D}Y(\mathfrak{g})}$ in the $\boldsymbol{RTT}$ formulation}\label{RTT}

Let $\mathfrak{g}$ be one of the classical Lie algebras $\mathfrak{gl}_N$, $\mathfrak{o}_{2n+1}$,
$\mathfrak{sp}_{2n}$ and $\mathfrak{o}_{2n}$ corresponding to $A$, $B$, $C$ and~$D$
classical series.
Let $N=2,3,\dots$ for $\mathfrak{gl}_N$,
$N=2n+1$ for $\mathfrak{g}=\mathfrak{o}_{2n+1}$ and
$N=2n$ for~$\mathfrak{g}=\mathfrak{sp}_{2n}$ or $\mathfrak{g}=\mathfrak{o}_{2n}$,
where $n=2,3,\dots$.

It is convenient to use positive and negative integers to index matrix elements of
operators from $\operatorname{End}\big({\CC}^{N}\big)$ for different algebras.
We introduce the sets of integers $\IA=1,2,\dots,N$; $\IB=-n,-n+1,\dots,-1,0,1,2,\dots,n$;
$\IC=\ID=-n+1,\dots,-1,0,1,2,\dots,n$. We will use notation~$\Ig$
to describe all sets of indices simultaneously.

\subsection[g-invariant R-matrix]{$\boldsymbol{\mathfrak{g}}$-invariant $\boldsymbol {R}$-matrix}

Let $R(u,v)$ be $\mathfrak{g}$-invariant $R$-matrix \cite{JLY18,ZZ79}
\begin{gather}\label{RBCD}
 R(u,v) = \mathbf{I}\otimes\mathbf{I} + \frac{c\, \mathbf{P}}{u-v} - \frac{c\,\mathbf{Q}}{u-v+c\kappa},
\end{gather}
where $\mathbf{I}=\sum_{i\in\Ig}\Ee_{i,i}$ is the identity operator acting in the space $\CC^{N}$ and
$\Ee_{i,j}\in\operatorname{End}\big(\CC^N\big)$ are $N\times N$ matrices with the only nonzero entry equals to 1 at
the intersection of the $i$-th row and~$j$-th column. The operators $\mathbf{P}$
and $\mathbf{Q}$ act in
$\CC^{N}\ot\CC^{N}$ such that
\begin{gather*}
\mathbf{P}=\sum_{i,j\in \Ig}\Ee_{i,j}\otimes\Ee_{j,i}, \qquad
\mathbf{Q}=\sum_{i,j\in \Ig}\epsilon_i\epsilon_j\Ee_{i,j}\otimes\Ee_{i',j'},
\end{gather*}
where
\begin{gather}\label{eps}
\epsilon_i=\begin{cases}
1&\text{for } \mathfrak{g}=\mathfrak{gl}_{N},\ \mathfrak{o}_{N},\\
- \sgn {(i - 1/2)} &\text{for } \mathfrak{g}=\mathfrak{sp}_{2n}
\end{cases}
\end{gather}
and
\begin{gather}\label{primemap}
i'=\begin{cases}
N+1-i&\text{for } \mathfrak{g}=\mathfrak{gl}_{N},\quad i\in \IA,\\
-i&\text{for } \mathfrak{g}=\mathfrak{o}_{2n+1},\quad i\in \IB,\\
-i+1&\text{for } \mathfrak{g}=\mathfrak{sp}_{2n},\mathfrak{o}_{2n},\quad i\in \IC,\ID.
\end{cases}
\end{gather}

Operators $\mathbf{P}$ and $\mathbf{Q}$ satisfy the properties
\begin{gather}\label{PQpro}
\mathbf{P}^2= \mathbf{I}\otimes\mathbf{I},\qquad \mathbf{Q}^2=N\mathbf{Q},\qquad
\mathbf{P}\cdot \mathbf{Q}=\mathbf{Q}\cdot \mathbf{P}=
\begin{cases}
\mathbf{Q}&\text{for } \mathfrak{g}=\mathfrak{o}_{N},\\
-\mathbf{Q}&\text{for } \mathfrak{g}=\mathfrak{sp}_{2n},
\end{cases}
\end{gather}
and parameter $\kappa$ in definition of $\mathfrak{g}$-invariant $R$-matrix \eqref{RBCD}
is equal to
\begin{gather}\label{kappa}
\kappa=
\begin{cases}
\infty \quad&\text{for } \mathfrak{g}=\mathfrak{gl}_{N},\\
N/2-1\quad&\text{for } \mathfrak{g}=\mathfrak{o}_{N},\\
n+1\quad&\text{for } \mathfrak{g}=\mathfrak{sp}_{2n}.
\end{cases}
\end{gather}
Value of $\kappa = \infty$ in case $\mathfrak{gl}_{N}$ means that one should drop the term with $Q$-operator in expression for $\mathfrak{gl}_N$-invariant $R$-matrix.
In \eqref{RBCD} nonzero $c\in\CC$ is a Yangian deformation parameter and~\mbox{$u,v\in\CC$} are spectral parameters.

Using properties \eqref{PQpro} one can check that $R$-matrix \eqref{RBCD} satisfies the Yang--Baxter
equation and unitarity condition
\begin{gather}\label{Runi}
R(u,v)\cdot R(v,u)=\sk{1-\frac{c^2}{(u-v)^2}}\mathbf{I}\otimes\mathbf{I}.
\end{gather}

\subsection[Yangian double DY(g)]{Yangian double $\boldsymbol{\DYg}$}

 In this paper we will use the Yangian doubles corresponding
to the classical Lia algebras $\mathfrak{g}$ of the series
$A$, $B$, $C$ and $D$ as it was introduced in the paper \cite{JLY18}.
$RTT$ realization of these algebras was denoted in \cite{JLY18} as $\mathcal{D}Y^R_c(\mathfrak{g})$.

The Yangian double $\mathcal{D}Y^R_c(\mathfrak{g})$
is an associative algebra generated by the elements
$T_{i,j}[\ell]$, $\ell\in\ZZ$ and $i,j\in \Ig$. These elements can be gathered into formal
series
\begin{gather*}
T^\pm_{i,j}(u)=\delta_{ij}+\sum_{\genfrac{}{}{0pt}{2}{\ell\geq0}{\ell<0}}\, T_{i,j}[\ell](u/c)^{-\ell-1},
\end{gather*}
which become the entries of the $T$-operators
\begin{gather*}
T^\pm(u)=\sum_{ i,j\in \Ig} \Ee_{ij} T^\pm_{i,j}(u)
\end{gather*}
satisfying the commutation relations\footnote{Let us recall, that we consider the Yangian double without central extension.}
\begin{gather}\label{rtt-dy}
 R(u,v) \left( T^\mu(u)\otimes\mathbf{I} \right) \left( \mathbf{I}\otimes T^\nu(v) \right) =
 \left( \mathbf{I}\otimes T^\nu(v) \right) \left( T^\mu(u)\otimes\mathbf{I} \right) R(u,v) ,
\end{gather}
where $\mu,\nu=\pm$.
The rational functions in $R$-matrix $R(u,v)$ should be understood as the series
expanded in powers $u^{-1}$
for the commutation relations with $\mu=+$ and $\nu=-$, e.g.,
\begin{gather*}
 \frac{1}{u-v+c\kappa}=\sum_{\ell=0}^\infty\frac{(v-c\kappa)^\ell}{u^{\ell+1}}.
\end{gather*}
On the other hand, these rational functions should be understood as the
series expanded in powers $v^{-1}$
for the commutation relations \eqref{rtt-dy} with $\mu=-$ and $\nu=+$.
Moreover, the diagonal
 entries of $T$-operators should be defined
 over $\CC[[c]]$ via $c$-adic topology. This ensures the invertibility
of the diagonal entries of $T$-operators. See the paper \cite{JLY18} for the
discussion on the topological aspects of the Yangian doubles
$\mathcal{D}Y^R_c(\mathfrak{g})$.
In our paper we used the notation $\DYg$ for the Yangian double
$\mathcal{D}Y^R_c(\mathfrak{g})$.

Sometimes we will call $T$-operators by {\it monodromy matrices} remembering that they become quantum
monodromies of some integrable model if $T_{i,j}[\ell]$ are
the operators acting in the Hilbert space of the corresponding physical model.

Equation \eqref{rtt-dy} yields the commutation relations of the monodromy matrix entries
\begin{gather}
 \big[ T^\mu_{i,j}(u), T^\nu_{k,l}(v) \big] =
 \frac{c}{u-v}\big( T^\nu_{k,j}(v)T^\mu_{i,l}(u) - T^\mu_{k,j}(u) T^\nu_{i,l}(v) \big)\nonumber
 \\ \hphantom{ \big[ T^\mu_{i,j}(u), T^\nu_{k,l}(v) \big] =}
{}+\frac{c}{u-v+c\kappa}\!\sum_{p\in \Ig}\epsilon_{p}\big(\delta_{k,i'}\,\epsilon_{i} \, T^\mu_{p,j}(u)T^\nu_{p',l}(v)-
 \delta_{l,j'}\,\epsilon_{j}\, T^\nu_{k,p'}(v)T^\mu_{i,p}(u)\big),\!
\label{rrt2}
\end{gather}
with $\epsilon_{i}$ defined in \eqref{eps}. Let us remind that the second line in \eqref{rrt2} will be missing
in case of the Yangian double $\DYGLN$.

For any matrix $M$ acting in $\CC^N$ we denote by $M^{\rm t}$ the transposition
\begin{gather*}
(M^{\rm t})_{i,j}=\epsilon_i \epsilon_j M_{j',i'}.
\end{gather*}
It is related to the ``usual'' transposition $(\cdot)^t$ by
a conjugation by the matrix $U=\sum_{i\in\Ig} \epsilon_i \Ee_{i,i'}$,
where $i'$ is defined in \eqref{primemap}. Obviously, $(M^{\rm t})^{\rm t}=M$.
Introduce the transpose-inverse $T$-operators
\begin{gather}\label{in-tr-mo}
\hTT^\pm(u)=\big(T^\pm(u)^{-1}\big)^{\rm t}.
\end{gather}
Possibility to inverse $T$-operators $T^\pm(u)^{-1}$ in $\DYg$ was discussed
in \cite{JLY18}.
One can check that transpose-inverse $T$-operator $\hTT^\pm(u)$
satisfy the same $RTT$ commutation relations \eqref{rrt2}.
To~prove it one needs to apply the transformation \eqref{in-tr-mo} it both spaces of the
$RTT$ relation \eqref{rtt-dy}.

The commutation relations for
the Yangian doubles $\DYBn$, $\DYCn$ and $\DYDn$ \eqref{rrt2},
imply the relations \cite{JLM18,JLY18}
\begin{gather}\label{qu-rel}
 \left( T^\pm(u-c\kappa)\right)^{\rm t} T^\pm(u)=
T^\pm(u) \left( T^\pm(u-c\kappa)\right)^{\rm t} = z^\pm(u) \mathbf{I},
\end{gather}
where $z^\pm(u)$ are central elements.
The $RTT$ algebras given by the commutation relations~\eqref{rrt2} with central elements $z^\pm(u)$ was denoted in [7] as
$\mathcal{D}X^R_c(\mathfrak{g})$.
Yangian doubles
$\DYBn$, $\DYCn$ and $\DYDn$ which we will consider in this paper are isomorphic to the
quotients of $\mathcal{D}X^R_c(\mathfrak{g})$ by the ideal generated by the modes of the
series $z^\pm(u)$. This means that we can set
 \begin{gather*}
z^\pm(u)=1
\end{gather*}
and the equality \eqref{qu-rel}
can be written in the form\footnote{From now on and until the end of the paper,
we assume that central elements $z^\pm(u)=1$ and we continue to use the notations
$T^\pm(u)$ and $\hat{T}^\pm(u)$ for the $T$-operators of the algebra $\DYg$.}
\begin{gather}\label{ident}
T^\pm_{i,j}(u-c\kappa)=\hTT^\pm_{i,j}(u),\qquad i,j\in \IB,\IC,\ID.
\end{gather}
It proves that a mapping
\begin{gather}\label{mapp}
T^\pm_{i,j}(u)\to \hat{T}^\pm_{i,j}(u)
\end{gather}
 is an automorphism of the $RTT$-algebra \eqref{rtt-dy}.
Some property of this automorphism is described in Appendix~\ref{ApB}.

The generating matrices $T^+(u)$ and $T^-(u)$ form two Borel subalgebras in the
algebras $\DYGLN$, $\DYBn$, $\DYCn$ and $\DYDn$. We denote each of these standard Borel subalgebras
as $\Ac^\pm$, $\Bc^\pm$, $\Cc^\pm$ and $\Dc^\pm$,
respectively.
Any of the Yangian doubles $\DYGLN$, $\DYBn$, $\DYCn$ and $\DYDn$
can be constructed by the quantum double construction starting from
one of its subalgebras \cite{Dr88}.
We denote these doubles as $\DAal$, $\DBal$,
 $\DCal$ and
 $\DDal$.
Quantum double construction in these cases uses the Hopf structure
\begin{gather}\label{stcop}
\Delta\big({T^\pm_{i,j}(u)}\big)=\sum_{\ell\in \Ig} T^\pm_{\ell,j}(u)\ot T^\pm_{i,\ell}(u),
\end{gather}
which shows that $T^+(u)$ and $T^-(u)$ generate also Hopf subalgebras in~$\DYg$.

\section[Gauss coordinates and the currents formulation of DY(g)]
{Gauss coordinates and the currents formulation of $\boldsymbol{\DYg}$}\label{GaussCoor}

As it is well known now, the infinite-dimensional algebras
which possess the $RTT$ formulation can be reformulated in terms of ``new'' or ``current'' realizations
 \cite{D88}.
An isomorphism between different realizations of the Yangian doubles and quantum affine algebras
was established first for the type $A$ algebras. The main
ingredients of this construction were the Gauss coordinates of $T$-operators.
Recently, it was discovered in the papers \cite{JLM18,JLM19,JLM19a,JLY18} that the same
mechanism allows to establish the corresponding isomorphisms between
$RTT$ and ``current'' realizations of the Yangian doubles and quantum affine algebras
for $B$, $C$ and $D$ series. In this section we des\-cribe these isomorphisms for each
of the algebras $\DYGLN$, $\DYBn$, $\DYCn$ and~$\DYDn$ separately.

\subsection[Gauss decomposition for DY(glN)]{Gauss decomposition for $\boldsymbol{\DYGLN}$}\label{3.1}

There are several different ways to introduce Gauss coordinates of $T$-operators. In this paper
we will use following Gauss decomposition of the monodromy matrix elements
\begin{gather}\label{GF}
\TT^\pm_{i,j}(u)=\sum_{\ell={\rm max}(i,j)}^{N} \FF^\pm_{\ell,i}(u)k^\pm_{\ell}(u)\EE^\pm_{j,\ell}(u),
\end{gather}
where we set $\FF^\pm_{i,i}(u)=\EE^\pm_{i,i}(u)=1$ and $\FF^\pm_{i,j}(u)=\EE^\pm_{j,i}(u)=0$ for $i<j$.
One can substitute~\eqref{GF} into commutation relations \eqref{rrt2} and check that Gauss coordinates
$\FF^\pm_{i+1,i}(u)$, $\EE^\pm_{i,i+1}(u)$, $i=1,\dots,N-1$ and $k^\pm_j(u)$, $j=1,\dots,N$
are generators of the Yangian double $\DYGLN$.

Let us introduce the rational function of the spectral parameters $u$ and $v$
\begin{gather}\label{fun}
 f(u,v)=\frac{u-v+c}{u-v}.
\end{gather}
The commutation relations between
generators of the Yangian double $\DYGLN$ can be rewritten in terms of so called {\it currents}
\begin{gather}\label{DF-iso1}
F_i({u})=\FF^{+}_{i+1,i}({u})-\FF^{-}_{i+1,i}({u}),\qquad
E_i({u})=\EE^{+}_{i,i+1}({u})-\EE^{-}_{i,i+1}({u})
\end{gather}
and can be presented in the form
(so called ``new'' realization of the Yangian double $\DYGLN$)
\begin{gather}
k^{\pm}_i(u) F_i(v) k^{\pm}_i(u)^{-1}=f(v,u)\,F_i(v),\nonumber
\\
k^{\pm}_{i+1}(u)F_i(v)k^{\pm}_{i+1}(u)^{-1}=f(u,v) F_i(v),\nonumber
\\
\label{tkiF}
k^{\pm}_i(u) F_j(v) k^{\pm}_i(u)^{-1}= {F_j(v),\qquad i\neq j,j+1,\qquad
1\leq j\leq N-1, \qquad 1\leq i\leq N};
\\
k^{\pm}_i(u)^{-1}E_i(v)k^{\pm}_i(u)=f(v,u) E_i(v),\nonumber
\\
k^{\pm}_{i+1}(u)^{-1}E_i(v)k^{\pm}_{i+1}(u)=f(u,v) E_i(v),\nonumber
\\
\label{tkiE}
k^{\pm}_i(u) E_j(v) k^{\pm}_i(u)^{-1}= E_j(v),\qquad i\neq j,j+1,\qquad
1\leq j\leq N-1, \qquad 1\leq i\leq N;
\\ \label{tFiFi}
f(u,v) F_i(u)F_i(v)= f(v,u) F_i(v)F_i(u);
\\ \label{tEiEi}
f(v,u) E_i(u) E_i(v)= f(u,v) E_i(v) E_i(u);
\\ \label{tFiFii}
(u-v-c) F_i(u)F_{i+1}(v)= (u-v) F_{i+1}(v)F_i(u);
\\ \label{tEiEii}
(u-v) E_i(u)E_{i+1}(v)= (u-v-c) E_{i+1}(v)E_i(u);
\\ \label{tEF}
[E_i(u),F_j(v)]=c\, \delta_{i,j}\, \delta(u,v)\big(k^+_{i}(u)\cdot k^+_{i+1}(u)^{-1}-k^-_{i}(v)\cdot k^-_{i+1}(v)^{-1}\big),
\end{gather}
and the Serre relations for the currents $E_i(u)$ and $F_i(u)$ (see, for example, \cite{HutLPRS17a}).
In \eqref{tEF} the symbol $\delta(u,v)$ means
the additive $\delta$-function given by the formal series
\begin{gather*}
\delta(u,v)=\frac{1}{u}\sum_{\ell\in\ZZ}\frac{v^\ell}{u^\ell}.
\end{gather*}

The main result of the paper \cite{LPRS19} was an explicit presentation of the isomorphism of the
Yangian double $\DYGLN$ \eqref{mapp} in terms of the Gauss coordinates.
Gauss decomposition of the monodromies $\hTT^\pm(u)$ has literally the same
form as in \eqref{GF} with the Gauss coordinates
$\FF^\pm_{j,i}(u)$, $\EE^\pm_{i,j}(u)$, $k^\pm_\ell(u)$ replaced by $\hFF^\pm_{j,i}(u)$, $\hEE^\pm_{i,j}(u)$, $\hk^\pm_\ell(u)$
such that for $1\leq i<j\leq N$ and $1\leq\ell\leq N$
\begin{gather}\label{hFF}
\hFF^\pm_{j,i}(u)=\tFF^\pm_{i',j'}(u-j'c),
\\
\label{hk}
\hk^\pm_{\ell}(u)=\frac{1}{k^\pm_{\ell'}(u-(\ell'-1)c)} \prod_{s=1}^{\ell'-1} \frac{k^\pm_s(u-s c)}{k^\pm_s(u-(s-1)c)} ,
\\
\label{hEE}
\hEE^\pm_{i,j}(u)=\tEE^\pm_{j',i'}(u-j'c),
\end{gather}
where (recall that indices $i'$, $j'$ and $\ell'$ are defined by \eqref{primemap})
\begin{gather}\label{tFF}
\tFF^\pm_{j,i}(u) =\sum_{\ell=0}^{j-i-1}(-)^{\ell+1}
\sum_{j>i_\ell>\cdots>i_1>i} \FF^\pm_{i_1,i}(u) \FF^\pm_{i_2,i_1}(u)\cdots \FF^\pm_{i_\ell,i_{\ell-1}}(u) \FF^\pm_{j,i_\ell}(u),
\\ \label{tEE}
\tEE^\pm_{i,j}(u) =\sum_{\ell=0}^{j-i-1}(-)^{\ell+1}
\sum_{j>i_\ell>\cdots>i_1>i} \EE^\pm_{i_\ell,j}(u) \EE^\pm_{i_{\ell-1},i_\ell}(u)\cdots \EE^\pm_{i_1,i_2}(u) \EE^\pm_{i,i_1}(u).
\end{gather}
The terms corresponding to the value $\ell=0$ in \eqref{tFF} and \eqref{tEE} are $-\FF_{j,i}(u)$ and
$-\EE_{i,j}(u)$ respectively.

Elements $\tFF^\pm_{j,i}(u)$ and $\tEE^\pm_{i,j}(u)$ given by \eqref{tFF} and \eqref{tEE} are matrix entries
of the inverse mat\-rices $\mathbf{F}^\pm(u)^{-1}$ and $\mathbf{E}^\pm(u)^{-1}$.
 Diagonal matrix
$\mathbf{K}^\pm(u)=\operatorname{diag}\big(k^\pm_1,\dots,k^\pm_N(u)\big)$, upper triangular matrix
$\mathbf{F}^\pm(u)$ and lower triangular matrix $\mathbf{E}^\pm(u)$
 define the Gauss decomposition
\eqref{GF} of the $T$-operators: $T^\pm(u)=\mathbf{F}^\pm(u)\cdot \mathbf{K}^\pm(u)\cdot \mathbf{E}^\pm(u)$.
Inverse $T$-operators
$T^\pm(u)^{-1}$ are equal to
\begin{gather}\label{inverse}
\tTT^\pm(u)=T^\pm(u)^{-1}=\mathbf{E}^\pm(u)^{-1}\cdot \mathbf{K}^\pm(u)^{-1}\cdot \mathbf{F}^\pm(u)^{-1}=
\tilde{\mathbf{E}}^\pm(u)\cdot \mathbf{K}^\pm(u)^{-1}\cdot
\tilde{\mathbf{F}}^\pm(u)
\end{gather}
and formulas \eqref{hFF}--\eqref{hEE} were proved in \cite{LPRS19} by reordering Gauss coordinates using the
commutation relations between them in the $T$-operators $\hTT^\pm(u)$ given by \eqref{in-tr-mo}.

In the current realization the Yangian double $\DYGLN$ possess the automorphism
\begin{gather}
F_i(u)\to \hF_i(u)=-F_{i'-1}(u-(i'-1)c),\qquad i=1,\dots,N-1 ,\nonumber
\\
E_i(u)\to \hE_i(u)=-E_{i'-1}(u-(i'-1)c),\qquad i=1,\dots,N-1 ,\nonumber
\\
k^\pm_\ell(u)\to \hk^\pm_\ell(u)=\frac{1}{k^\pm_{\ell'}(u-(\ell'-1)c)}
\prod_{s=1}^{\ell'-1} \frac{k^\pm_s(u-s c)}{k^\pm_s(u-(s-1)c)},\qquad \ell=1,\dots,N,
\label{map-cur}
\end{gather}
which is induced by \eqref{hFF}--\eqref{hEE}. It can be directly verified
 using the commutation relations
 \eqref{tkiF}--\eqref{tEF}.

In the next three subsections we will introduce Gauss coordinates for the $T$-operators and the corresponding currents
 for the Yangian doubles $\DYBn$, $\DYCn$ and $\DYDn$. We~will use the same notations for
 these quantities in each of these subsections, but
 it will be always clear from the context what
 Gauss coordinates, currents and $T$-operators we are considering.

\subsection[Gauss coordinates for DY(o2n+1)]{Gauss coordinates for $\boldsymbol{\DYBn}$}\label{3.2}
Let $e_1,\dots,e_{n}$ be an orthonormal basis in the $n$-dimensional
Euclidean space. The simple roots~$\alpha_i$ for $i=0,1,\dots,n-1$ of the algebra $\mathfrak{o}_{2n+1}$ are
\begin{gather*}
\alpha_0=e_1,\qquad \alpha_i=e_{i+1}-e_i,\qquad i=1,\dots,n-1.
\end{gather*}
We introduce Gauss
coordinates for the monodromy matrix $T^\pm(u)\in\DYBn$ similarly to~\eqref{GF}
\begin{gather}\label{GaussB}
T^\pm_{i,j}(u)={ \sum_{{\ell = \rm max}(i,j)}^{n} } \FF^\pm_{\ell,i}(u)k^\pm_\ell(u)\EE^\pm_{j,\ell}(u),
\end{gather}
where we again assume that $\FF^\pm_{i,j}(u)=\EE^\pm_{j,i}(u)=0$ for $i<j$ and $\FF^\pm_{i,i}(u)=\EE^\pm_{i,i}(u)=1$ for
$i,j\in\IB$.

The commutation relations for the Yangian double $\DYBn$ \eqref{rrt2}
for the monodromy matrix entries
 $T^\pm_{i,j}(u)$ for
$-n\leq i,j\leq 0$ and $0\leq i,j\leq n$ satisfy the
$\mathfrak{gl}_{n+1}$-type commutation
relations except the relations between
$T^\pm_{i,j}(u)$ and $T^\pm_{k,l}(v)$ when either
$i=k=0$ or $j=l=0$ or~$i=k=n=l=0$.
Using \cite{LPRS19} we can express part of
the Gauss coordinates of the transpose-inverse
monodromy matrices $\hTT^\pm(u)$ given by \eqref{in-tr-mo}
through Gauss coordinates of $T^\pm(u)$ \eqref{GaussB}
($0\leq i<j\leq n$)
\begin{gather}
\label{hFFodd}
\hFF^\pm_{j,i}(u)=\tFF^\pm_{i',j'}(u-c(n-j+1)),
\\
\label{hkodd}
\hk^\pm_{j}(u)=\frac{1}{k^\pm_{j'}(u-c(n-j))}
\prod_{\ell=j+1}^{n}
\frac{k^\pm_{\ell'}(u-c(n-\ell+1))}{k^\pm_{\ell'}(u-c(n-\ell))} ,
\\
\label{hEEodd}
\hEE^\pm_{i,j}(u)=\tEE^\pm_{j',i'}(u-c(n-j+1)).
\end{gather}
Recall that $i'=-i$, $j'=-j$ and $\ell'=-\ell$ in this case.
Formulas \eqref{hFFodd}--\eqref{hEEodd} can be obtained from the corresponding
formulas \eqref{hFF}--\eqref{hEE} if one sets there $N=2n+1$ and uses overall shift in the indices
$i=\tilde{i}+n+1$ and $j=\tilde{j}+n+1$ for $-n\leq\tilde{i},\tilde{j}\leq n$.

The Gauss coordinates $\tFF^\pm_{i,j}(u)$ in \eqref{hFFodd} and
$\tEE^\pm_{i,j}(u)$ in \eqref{hEEodd} are given by the formulas \eqref{tFF} and \eqref{tEE}.
To obtain \eqref{hFFodd} we have used only
$\mathfrak{gl}_{n+1}$-type commutation between $T$-operator entries. It explains
 why these formulas do not differ from the formulas
\eqref{hFF}--\eqref{hEE} up to certain shift in the spectral parameter.

Identification \eqref{ident} of the matrix entries
$T^\pm_{i,j}(u-c\kappa)=\hTT^\pm_{i,j}(u)$ for $0\leq i,j\leq n$
 allows to identify
 the Gauss coordinates corresponding to the simple roots
 $\pm\alpha_i$, $i=0,\dots,n-1$ of the algebra $\mathfrak{o}_{2n+1}$
 (recall that $\kappa=n-1/2$)
\begin{gather}
\label{F-id}
\FF^\pm_{i+1,i}(u)=-\FF^\pm_{-i,-i-1}(u+c(i-1/2)),
\\
\label{E-id}
\EE^\pm_{i,i+1}(u)=-\EE^\pm_{-i-1,-i}(u+c(i-1/2))
\end{gather}
for $0\leq i\leq n-1$ and for $0\leq \ell\leq n$
\begin{gather}
\label{k-id}
k^\pm_\ell(u)=\frac{1}{k^\pm_{-\ell}(u+c(\ell-1/2))}\prod_{s=\ell+1}^{n}
\frac{k^\pm_{-s}(u+c(s-3/2))}{k^\pm_{-s}(u+c(s-1/2))}.
\end{gather}
As usual, we associate the Gauss coordinates $\FF^\pm_{i+1,i}(u)$ and
$\FF^\pm_{-i,-i-1}(u)$ with the negative simple roots $-\alpha_i$,
and the Gauss coordinates $\EE^\pm_{i,i+1}(u)$ and
$\EE^\pm_{-i-1,-i}(u)$ with the positive simple roots~$\alpha_i$.

Formulas \eqref{F-id}--\eqref{k-id} can be inverted to express $\FF^\pm_{i+1,i}(u)$, $\EE^\pm_{i,i+1}(u)$ for $-n\leq i\leq -1$ and~$k^\pm_\ell(u)$ for $-n\leq\ell\leq 0$
through the Gauss coordinates $\FF^\pm_{i+1,i}(u)$, $\EE^\pm_{i,i+1}(u)$ for $0\leq i\leq n-1$ and~$k^\pm_\ell(u)$ for $0\leq\ell\leq n$. They are
\begin{gather}
\FF^\pm_{-i,-i-1}(u)=-\FF^\pm_{i+1,i}(u-c(i-1/2)),\qquad 0\leq i\leq n-1,\nonumber
\\
\EE^\pm_{-i-1,-i}(u)=-\EE^\pm_{i,i+1}(u-c(i-1/2)),\qquad 0\leq i\leq n-1,\nonumber
\\
k^\pm_{-\ell}(u)=\frac{1}{k^\pm_{\ell}(u-c(\ell-1/2))}\prod_{s=\ell+1}^{n}
\frac{k^\pm_{s}(u-c(s-3/2))}{k^\pm_{s}(u-c(s-1/2))}.
\label{oth-ge}
\end{gather}

Equalities \eqref{F-id}--\eqref{k-id} and \eqref{oth-ge} allow to chose the set of independent generators
of subalgebras $\Bal^\pm$. This can be either the set
\begin{gather}\label{choice1}
\FF^\pm_{i+1,i}(u),\qquad \EE^\pm_{i,i+1}(u),\qquad
0\leq i\leq n-1,\qquad k^\pm_j(u),\qquad {1}\leq j\leq n,
\end{gather}
or
\begin{gather*}
\FF^\pm_{i+1,i}(u),\qquad \EE^\pm_{i,i+1}(u),\qquad -n\leq i\leq -1,\qquad k^\pm_j(u),\qquad -n\leq j\leq {-1.}
\end{gather*}
The modes of the currents $k^\pm_0(u)$ which enters the commutation relations \eqref{tkiFB}, \eqref{tkiEB} and~\eqref{tEFB} can be obtained from the relation
\begin{gather}
\label{restrel}
k^\pm_{0}(u+c/2)k^\pm_{0}(u)=\prod_{s=1}^{n}\frac{k^\pm_s(u-c(s-3/2))}{k^\pm_s(u-c(s-1/2))}.
\end{gather}
In what follows we will use the set \eqref{choice1} as the set of generators of the algebra $\DBal$.

Besides rational function \eqref{fun} we introduce the function
\begin{gather}\label{fgo}
 \fgo(u,v)=\frac{u-v+c/2}{u-v}.
\end{gather}
In order to find current realization of the algebra $\DYBn$ and due to \eqref{ident}
it is enough to~consider the
commutation relations for the monodromy matrix elements $T_{i,j}(u)$ following from~\eqref{rtt-dy}
for $0\leq i,j\leq n$ only. Then the commutation relations in $\DBal$
can be written in terms of the Cartan currents $k^\pm_j(u)$ for $0\leq j\leq n$
and generating series (currents) associated to the
 simple roots $\pm\alpha_i$
of the algebra $\mathfrak{o}_{2n+1}$
\begin{gather}\label{DF-iso1B}
F_i({u})=\FF^{+}_{i+1,i}({u})-\FF^{-}_{i+1,i}({u}),\qquad
E_i({u})=\EE^{+}_{i,i+1}({u})-\EE^{-}_{i,i+1}({u}),
\end{gather}
for $0\leq i\leq n-1$ as follows \cite{JLY18}
\begin{gather}
k^{\pm}_{0}(u)F_{0}(v)k^{\pm}_{0}(u)^{-1}=f(u,v)f(v,u+c/2)F_{0}(v),\nonumber
\\
k^{\pm}_i(u) F_i(v) k^{\pm}_i(u)^{-1}= f(v,u) F_i(v),\qquad 1\leq i\leq n-1,\nonumber
\\
k^{\pm}_{i+1}(u)F_i(v)k^{\pm}_{i+1}(u)^{-1}= f(u,v) F_i(v),\qquad 0\leq i\leq n-1,\nonumber
\\
k^{\pm}_i(u) F_j(v) k^{\pm}_i(u)^{-1}= F_j(v),\qquad i\neq j,j+1,\qquad
0\leq j\leq n-1, \qquad 0\leq i\leq n;
\label{tkiFB}
\\
k^{\pm}_{0}(u)^{-1}E_{0}(v)k^{\pm}_{0}(u)=f(u,v)f(v,u+c/2)E_{0}(v),\nonumber
\\
k^{\pm}_i(u)^{-1}E_i(v)k^{\pm}_i(u)=f(v,u) E_i(v),\qquad 1\leq i\leq n-1,\nonumber
\\
k^{\pm}_{i+1}(u)^{-1}E_i(v)k^{\pm}_{i+1}(u)=f(u,v) E_i(v),\qquad 0\leq i\leq n-1,\nonumber
\\
{k^{\pm}_i(u) E_j(v) k^{\pm}_i(u)^{-1}}= E_j(v),\qquad i\neq j,j+1,\qquad
0\leq j\leq n-1, \qquad 0\leq i\leq n;
\label{tkiEB}
\\
\fgo(u,v) F_{0}(u)F_{0}(v)= \fgo(v,u) F_{0}(v)F_{0}(u),\nonumber
\\
f(u,v) F_i(u)F_i(v)= f(v,u) F_i(v)F_i(u),\qquad 1\leq i\leq n-1;
\label{tFiFiB}
\\
\fgo(v,u) E_{0}(u)E_{0}(v)= \fgo(u,v) E_{0}(v)E_{0}(u),\nonumber
\\
f(v,u) E_i(u) E_i(v)= f(u,v) E_i(v) E_i(u),\qquad 1\leq i\leq n-1;
\label{tEiEiB}
\\
\label{tFiFiiB}
(u-v-c) F_i(u)F_{i+1}(v)= (u-v) F_{i+1}(v)F_i(u),\qquad 0\leq i\leq n-2;
\\
\label{tEiEiiB}
(u-v) E_i(u)E_{i+1}(v)= (u-v-c) E_{i+1}(v)E_i(u),\qquad 0\leq i\leq n-2;
\\
\label{tEFB}
[E_i(u),F_j(v)]=c\, \delta_{i,j}\, \delta(u,v)\big(k^+_{i}(u)\cdot
k^+_{i+1}(u)^{-1}-k^-_{i}(v)\cdot k^-_{i+1}(v)^{-1}\big),
\end{gather}
and the Serre relations for the currents $E_i(u)$ and $F_i(u)$, see, e.g.,~\cite{JLY18}.

As in the case of $A$-type algebras all
equalities in \eqref{tkiFB}--\eqref{tEFB} should be understood as equalities between formal series,
so that they correspond to
a countable number of relations between modes of the currents. The proof that these relations
follow from the $RTT$ commutation relations \eqref{rrt2} is a straightforward repetition of the arguments invented
in \cite{DF93} and exploited in the present situation in \cite{JLM18,JLY18}.

\subsection[Gauss coordinates for DY(sp2n)]{Gauss coordinates for $\boldsymbol{\DYCn}$}

The simple roots $\alpha_i$, $0\leq i\leq n-1$ for the algebra $\mathfrak{sp}_{2n}$ are
\begin{gather*}
\alpha_0=2e_1,\qquad \alpha_i=e_{i+1}-e_i,\qquad i=1,\dots,n-1.
\end{gather*}
For the Yangian double $\DYCn$ we repeat the same approach that was realized in the previous section.
We use the Gauss decomposition \eqref{GaussB} where now $i,j\in\IC$.
Due to \eqref{rrt2} the mono\-dromy matrix
elements $T^\pm_{i,j}(u)$ for the values of the indices $-n+1\leq i,j\leq 0$ and
$1\leq i,j\leq n$ satisfy the commutation relations of the Yangian double $\DYAn$.
Repeating calculations of the previous section
we can express part of the Gauss coordinates
of the inverse-transpose monodromy matrix elements
$\hTT^\pm_{i,j}(u)$ \eqref{in-tr-mo} through the Gauss coordinates
of $T^\pm(u)\in\DYCn$. These relations are given by the formulas
 \eqref{hFFodd}--\eqref{hEEodd},
where $i'=-i+1$, $j'=-j+1$ and $\ell'=-\ell+1$. Then the identification \eqref{ident} for
$1\leq i,j\leq n$ with $\kappa=n+1$ yields the identification of the Gauss
coordinates corresponding to the
simple roots $\pm\alpha_i$,
$1\leq i\leq n-1$
for the algebra~$\mathfrak{sp}_{2n}$
\begin{gather}
\FF^\pm_{-i+1,-i}(u)=-\FF^\pm_{i+1,i}(u-c(i+1)),\qquad 1\leq i\leq n-1,\nonumber
\\
\EE^\pm_{-i,-i+1}(u)=-\EE^\pm_{i,i+1}(u-c(i+1)),\qquad 1\leq i\leq n-1,\nonumber
\\
k^\pm_{-\ell+1}(u)=\frac{1}{k^\pm_{\ell}(u-c(\ell+1))}\prod_{s=\ell+1}^{n}
\frac{k^\pm_{s}(u-c s)}{k^\pm_{s}(u-c(s+1))},
\label{sp-ge}
\end{gather}
where $\ell=1,\dots,n$.

Equalities \eqref{sp-ge} allow to chose the set of independent generators
of subalgebras $\Cal^\pm$. We~cho\-ose the set of the Gauss coordinates
\begin{gather}\label{ch1C}
\FF^\pm_{i+1,i}(u),\qquad \EE^\pm_{i,i+1}(u),\qquad 0\leq i\leq n-1,\qquad
k^\pm_j(u),\qquad 1\leq j\leq n,
\end{gather}
and this will be a set of independent generators of the Yangian double $\DCal$.
Note, that the set~\eqref{ch1C} includes also Gauss coordinates $\FF^\pm_{1,0}(u)$ and
$\EE^\pm_{0,1}(u)$ which correspond to the simple roots~$\pm\alpha_0$.

To describe the current realization of the Yangian double $\DCal$ we replace function $\fgo(u,v)$
introduced by \eqref{fgo} by the function
\begin{gather*}
 \fgo(u,v)=\frac{u-v+2c}{u-v}.
\end{gather*}
To find current realization of the algebra $\DYCn$ it is enough to consider the
commutation relations for the monodromy matrix elements $T_{i,j}(u)$ following from \eqref{rrt2}
for $0\leq i,j\leq n$ only. Then the commutation relations in $\DCal$
can be written in terms of the Cartan currents $k^\pm_\ell(u)$ for $1\leq \ell\leq n$
and generating series (currents) given by the Ding--Frenkel formulas \eqref{DF-iso1B}
as follows~\cite{JLY18}
\begin{gather}
k^{\pm}_{1}(u)F_{0}(v)k^{\pm}_{1}(u)^{-1}=\fgo(u,v) F_{0}(v),\nonumber
\\
k^{\pm}_i(u) F_i(v) k^{\pm}_i(u)^{-1}= f(v,u) F_i(v),\qquad 1\leq i\leq n-1,\nonumber
\\
k^{\pm}_{i+1}(u)F_i(v)k^{\pm}_{i+1}(u)^{-1}= f(u,v) F_i(v),\qquad 1\leq i\leq n-1,\nonumber
\\
{k^{\pm}_i(u) F_j(v) k^{\pm}_i(u)^{-1}}= F_j(v),\qquad \forall\, i\neq j,j+1,\qquad
0\leq j\leq n-1;
\label{tkiFC}
\\
k^{\pm}_{1}(u)^{-1}E_{0}(v)k^{\pm}_{1}(u)=\fgo(u,v)E_{0}(v),\nonumber
\\
k^{\pm}_i(u)^{-1}E_i(v)k^{\pm}_i(u)=f(v,u) E_i(v),\qquad 1\leq i\leq n-1,\nonumber
\\
k^{\pm}_{i+1}(u)^{-1}E_i(v)k^{\pm}_{i+1}(u)=f(u,v) E_i(v),\qquad 1\leq i\leq n-1,\nonumber
\\
{k^{\pm}_i(u) E_j(v) k^{\pm}_i(u)^{-1}}= E_j(v),\qquad \forall\, i\neq j,j+1,\qquad
0\leq j\leq n-1;
\label{tkiEC}
\\
\fgo(u,v) F_{0}(u)F_{0}(v)= \fgo(v,u) F_{0}(v)F_{0}(u),\nonumber
\\
f(u,v) F_i(u)F_i(v)= f(v,u) F_i(v)F_i(u),\qquad 1\leq i\leq n-1;\label{tFiFiC}
\\
\fgo(v,u) E_{0}(u)E_{0}(v)= \fgo(u,v) E_{0}(v)E_{0}(u),\nonumber
\\
f(v,u) E_i(u) E_i(v)= f(u,v) E_i(v) E_i(u),\qquad 1\leq i\leq n-1;
\label{tEiEiC}
\\
(u-v-2c) F_0(u)F_{1}(v)= (u-v) F_{1}(v)F_0(u),\nonumber
\\
(u-v-c) F_i(u)F_{i+1}(v)= (u-v) F_{i+1}(v)F_i(u),\qquad 1\leq i\leq n-2;
\label{tFiFiiC}
\\
(u-v) E_0(u)E_{1}(v)= (u-v-2c) E_{1}(v)E_0(u),\nonumber
\\
(u-v) E_i(u)E_{i+1}(v)= (u-v-c) E_{i+1}(v)E_i(u),\qquad 1\leq i\leq n-2;
\label{tEiEiiC}
\\
\label{tEFC}
[E_i(u),F_j(v)]=c\, (1+\delta_{i,0})\, \delta_{i,j}\, \delta(u,v)\big(k^+_{i}(u)\cdot k^+_{i+1}(u)^{-1}-k^-_{i}(v)\cdot k^-_{i+1}(v)^{-1}\big)
\end{gather}
and the Serre relations for the currents $E_i(u)$ and $F_i(u)$, $0\leq i\leq n-1$, see, e.g., \cite{JLY18}.
In \eqref{tEFC} the Cartan currents $k^\pm_0(u)$ are given by the last relation in \eqref{sp-ge}
for $\ell=1$
\begin{gather*}
k^\pm_{0}(u)=\frac{1}{k^\pm_{1}(u-2c)}\prod_{s=2}^{n}
\frac{k^\pm_{s}(u-c s)}{k^\pm_{s}(u-c(s+1))}.
\end{gather*}
For the proof of these relations we address readers to the paper \cite{JLY18}.

\subsection[Gauss coordinates for DY(o2n)]{Gauss coordinates for $\boldsymbol{\DYDn}$}\label{3.4}

The simple roots $\alpha_i$, $0\leq i\leq n-1$ for the algebra $\mathfrak{o}_{2n}$ are
\begin{gather*}
\alpha_0=e_1+e_2,\qquad
\alpha_i=e_{i+1}-e_i,\qquad
i=1,\dots,n-1.
\end{gather*}
For the Yangian double $\DYDn$ we follow the same approach.
We use the Gauss decomposition~\eqref{GaussB} with indices $i,j\in\ID$
using properties of the Gauss coordinates formulated in this case as following
\begin{prop}
\begin{gather}
\label{Drest}
\FF^\pm_{1,0}(u)=\EE^\pm_{0,1}(u)=0,
\\
\label{Dprop1}
\FF^\pm_{2,0}(u)=-\FF^\pm_{1,-1}(u),\qquad \EE^\pm_{0,2}(u)=-\EE^\pm_{-1,1}(u),
\\
\label{Dprop2}
 [\FF^\mu_{2,0}(u),\FF^\nu_{2,1}(v)]=0,\qquad
 [\EE^\mu_{0,2}(u),\EE^\nu_{1,2}(v)]=0,\qquad\mu,\nu=\pm.
\end{gather}
\end{prop}

This proposition was proved in the papers \cite{JLM18,JLY18}. In particular, restrictions \eqref{Drest}
were mentioned in Proposition~5.11 of~\cite{JLM18} and Proposition~5.18 of~\cite{JLY18};
the equalities \eqref{Dprop1} were proved in Proposition~5.7 of~\cite{JLM18};
commutativity of the Gauss coordinates \eqref{Dprop2} follows from the commutativity \eqref{tFiFiiD} of the
corresponding currents proved in Theorem~5.22 of~\cite{JLY18}.

As well as in the previous section the monodromy matrix
elements $T^\pm_{i,j}(u)$ for the values of the indices $-n+1\leq i,j\leq 0$ and
$1\leq i,j\leq n$ satisfy the commutation relations of the Yangian double $\DYAn$.
 Gauss coordinates
of the inverse-transpose monodromy matrix
$\hTT^\pm(u)$ \eqref{in-tr-mo} are related to the Gauss coordinates of $T^\pm(u)\in\DYDn$ by the
equalities \eqref{hFFodd}--\eqref{hEEodd}. Equations~\eqref{ident} for
$1\leq i,j\leq n$ with $\kappa=n-1$ allows to obtain the identification of the Gauss
coordinates corresponding to the
simple roots $\pm\alpha_i$,
$1\leq i\leq n-1$
for the algebra $\mathfrak{o}_{2n}$
\begin{gather}
\FF^\pm_{-i+1,-i}(u)=-\FF^\pm_{i+1,i}(u-c(i-1)),\qquad 1\leq i\leq n-1,\nonumber
\\
\EE^\pm_{-i,-i+1}(u)=-\EE^\pm_{i,i+1}(u-c(i-1)),\qquad 1\leq i\leq n-1,\nonumber
\\
k^\pm_{-\ell+1}(u)=\frac{1}{k^\pm_{\ell}(u-c(\ell-1))}\prod_{s=\ell+1}^{n}
\frac{k^\pm_{s}(u-c(s-2))}{k^\pm_{s}(u-c(s-1))},
\label{o2n-ge}
\end{gather}
where $\ell=1,\dots,n$.
Note that formulas \eqref{o2n-ge} for identification of the Gauss coordinates in the Yangian double
$\DYDn$ differ from the formulas \eqref{sp-ge} for the analogous identification in the Yangian double
$\DYCn$ by the overall shift by $2c$ in the right hand sides of \eqref{o2n-ge}. This is because
the method of identification of the Gauss coordinates is the same for these algebras, but $\kappa=n+1$ for
 $\DYCn$ and $\kappa=n-1$ for $\DYDn$.

 Due to the identification formulas \eqref{Dprop1},
 \eqref{o2n-ge} and the commutation relations \eqref{rrt2},
 the independent set of generators of the Yangian double $\DYDn$ is the collection of the
 Gauss coordinates
 \begin{gather}
 \FF^\pm_{2,0}(u),\qquad \EE^\pm_{0,2}(u),\qquad \FF^\pm_{i+1,i}(u),\qquad \EE^\pm_{i,i+1}(u),\qquad
 i=1,\dots,n-1,\nonumber
 \\
 k^\pm_j(u),\qquad j=1,\dots,n.
 \label{ch1D}
 \end{gather}
 The Gauss coordinates
 $\FF^\pm_{2,0}(u)$ and $\EE^\pm_{0,2}(u)$ correspond to the simple roots $\mp\alpha_0$
 \cite{JLY18}.

The currents for the ``new'' realization of the Yangian double $\DYDn$ are defined by the
standard Ding--Frenkel formulas
\begin{gather*}
F_0(u)=\FF^+_{2,0}(u)-\FF^-_{2,0}(u),\qquad F_i(u)=\FF^+_{i+1,i}(u)-\FF^-_{i+1,i}(u),
\\
E_0(u)=\EE^+_{0,2}(u)-\EE^-_{0,2}(u),\qquad E_i(u)=\EE^+_{i,i+1}(u)-\EE^-_{i,i+1}(u)
\end{gather*}
and they satisfy the commutation relations
\begin{gather}
k^{\pm}_{1}(u)F_{0}(v)k^{\pm}_{1}(u)^{-1}=f(u,v) F_{0}(v),\nonumber
\\
k^{\pm}_{2}(u)F_{0}(v)k^{\pm}_{2}(u)^{-1}=f(u,v) F_{0}(v),\nonumber
\\
k^{\pm}_{j}(u)F_{0}(v)k^{\pm}_{j}(u)^{-1}=F_{0}(v),\qquad 3\leq j\leq n,\nonumber
\\
k^{\pm}_i(u) F_i(v) k^{\pm}_i(u)^{-1}= f(v,u) F_i(v),\qquad 1\leq i\leq n-1,\nonumber
\\
k^{\pm}_{i+1}(u)F_i(v)k^{\pm}_{i+1}(u)^{-1}= f(u,v) F_i(v),\qquad 1\leq i\leq n-1,\nonumber
\\
{k^{\pm}_i(u) F_j(v) k^{\pm}_i(u)^{-1}}= F_j(v),\qquad \forall\, i\neq j,j+1,\qquad
1\leq j\leq n-1;\nonumber
\\
k^{\pm}_{1}(u)^{-1}E_{0}(v)k^{\pm}_{1}(u)=f(u,v)E_{0}(v),\nonumber
\\
k^{\pm}_{2}(u)^{-1}E_{0}(v)k^{\pm}_{2}(u)=f(u,v)E_{0}(v),\nonumber
\\
k^{\pm}_{j}(u)^{-1}E_{0}(v)k^{\pm}_{j}(u)=E_{0}(v),\qquad 3\leq j\leq n,\nonumber
\\
k^{\pm}_i(u)^{-1}E_i(v)k^{\pm}_i(u)=f(v,u) E_i(v),\qquad 1\leq i\leq n-1,\nonumber
\\
k^{\pm}_{i+1}(u)^{-1}E_i(v)k^{\pm}_{i+1}(u)=f(u,v) E_i(v),\qquad 1\leq i\leq n-1,\nonumber
\\
{k^{\pm}_i(u) E_j(v) k^{\pm}_i(u)^{-1}}= E_j(v),\qquad \forall\, i\neq j,j+1,\qquad
1\leq j\leq n-1;\nonumber
\\
f(u,v) F_i(u)F_i(v)= f(v,u) F_i(v)F_i(u),\qquad 0\leq i\leq n-1;\nonumber
\\
f(v,u) E_i(u) E_i(v)= f(u,v) E_i(v) E_i(u),\qquad 0\leq i\leq n-1;\nonumber
\\
F_0(u)F_{1}(v)= F_{1}(v)F_0(u),\nonumber
\\
(u-v-c) F_0(u)F_{2}(v)= (u-v) F_{2}(v)F_0(u),\nonumber
\\
(u-v-c) F_i(u)F_{i+1}(v)= (u-v) F_{i+1}(v)F_i(u),\qquad 1\leq i\leq n-2;
\label{tFiFiiD}
\\
 E_0(u)E_{1}(v)= E_{1}(v)E_0(u),\nonumber
 \\
(u-v) E_0(u)E_{2}(v)= (u-v-c) E_{2}(v)E_0(u),\nonumber
\\
(u-v) E_i(u)E_{i+1}(v)= (u-v-c) E_{i+1}(v)E_i(u),\qquad 1\leq i\leq n-2;\nonumber
\\
\label{tEFD}
[E_i(u),F_j(v)]=c\, \delta_{i,j}\, \delta(u,v)\big(k^+_{i}(u)\cdot k^+_{i+1+\delta_{i,0}}(u)^{-1}-
k^-_{i}(v)\cdot k^-_{i+1+\delta_{i,0}}(v)^{-1}\big)
\end{gather}
and the Serre relations for the currents $E_i(u)$ and $F_i(u)$, $0\leq i\leq n-1$ \cite{JLY18}.
Cartan currents~$k^\pm_0(u)$ in \eqref{tEFD} are given by the last equality in \eqref{o2n-ge} for $\ell=1$
\begin{gather*}
k^\pm_{0}(u)=\frac{1}{k^\pm_{1}(u)}\prod_{\ell=2}^{n}
\frac{k^\pm_{\ell}(u-c(\ell-2))}{k^\pm_{\ell}(u-c(\ell-1))}.
\end{gather*}

\section{Gauss coordinates and projections}\label{GCandproj}

Before we proceed to the relations between Gauss coordinates and the currents
we have to describe the projections onto intersections of the different type Borel subalgebras
in the Yangian doubles. The mathematically rigorous definition of these projections
for the quantum affine algebras was given in the paper \cite{EKhP07}. For the Yangian doubles
they can be defined analogously (see, for example, \cite{HutLPRS17a}).
In this paper we will use more practical definition of these projections which will be described
at the end of this section.

Description of the different types Borel subalgebras is very similar for all types Yangian doubles
$\DYg$, where $\mathfrak{g}=\mathfrak{gl}_N,\mathfrak{o}_{2n+1}, \mathfrak{sp}_{2n},\mathfrak{o}_{2n}$.
The only difference will be in the sets of~lgebraically independent sets of generators for each
algebra. That is why we will use a notation~$\Xc^\pm$ for the standard Borel subalgebras
in each Yangian double $\DYg$. Namely, $\Xc^\pm=\Ac^\pm$ for $\DYGLN$,
 $\Xc^\pm=\Bc^\pm$ for $\DYBn$, $\Xc^\pm=\Cc^\pm$ for $\DYCn$ and
 $\Xc^\pm=\Dc^\pm$ for $\DYDn$. In all cases subalgebras $\Xc^\pm$ are
 generated by the entries of
$T$-operators $T^\pm_{i,j}(u)$. These subalgebras
are Hopf subalgebras with respect to the coproduct \eqref{stcop} and the commutation relations are given
by the relations \eqref{rrt2}. The whole Yangian double $\DYg$ may be obtained
by the quantum double construction \cite{Dr88} from any of its Borel subalgebra $\Xc^+$ or
$\Xc^-$ using these algebraic and coalgebraic structures.

The ``current'' realizations of the Yangian doubles can be also obtained by the quantum double
construction, but in this case one has to chose another type Borel subalgebras and
another coalgebraic structure.

Let $\Xc_f^\pm$, $\Xc_e^\pm$ and $\Xc_k^\pm$ are subalgebras in the Yangian double $\DYg$
generated by the nonnegative and negative modes of the simple root currents
$F_i(u)$, $E_i(u)$, $1\leq i\leq N-1$
and~$k^\pm_j(u)$, $1\leq j\leq N$ for $\DYGLN$; $F_i(u)$, $E_i(u)$, $0\leq i\leq n-1$
and $k^\pm_j(u)$, $1\leq j\leq n$ for~$\DYBn$, $\DYCn$, $\DYDn$ with additional
currents $k^\pm_0(u)$ for $\DYBn$ which satisfy the relations~\eqref{restrel}.
It is clear that standard Borel subalgebras
$\Xc^\pm$ are $\Xc_f^\pm\cup \Xc_k^\pm\cup \Xc_e^\pm$.

Different choice of the initial Borel sualgebras in the quantum double construction
of $\DYg$
is to consider $\Xc_F=\Xc_f\cup\Xc_{k}^+$ as the union of the two subalgebras
$\Xc_f$ and $\Xc_{k}^+$ formed by all modes of the simple root currents $F_i(u)$ and modes
of the ``positive'' Cartan currents $k^+_j(u)$
\begin{gather}\label{modcF}
F_i(u)=\sum_{\ell\in\ZZ}F_i[\ell]u^{-\ell-1},\qquad k^+_j(u)=1+\sum_{\ell\geq 0}k_j[\ell]u^{-\ell-1}.
\end{gather}
We call this type Borel subalgebra as the {\it current} Borel subalgebra.
The other or dual current subalgebra $\Xc_E=\Xc_e\cup\Xc_{k}^-$ is formed by all modes of the
simple root currents $E_i(u)$ and modes of the ``negative'' Cartan currents $k^-_j(u)$
\begin{gather}\label{modcE}
E_i(u)=\sum_{\ell\in\ZZ}E_i[\ell]u^{-\ell-1},\qquad k^-_j(u)=1-\sum_{\ell<0}k_j[\ell]u^{-\ell-1}.
\end{gather}
The fact that $\Xc_F$ and $\Xc_E$ are subalgebras in the Yangian double $\DYg$
is clearly seen from the commutation relations in terms of the currents given in the Sections~\ref{3.1}--\ref{3.4}.

Current Borel subalgebras $\Xc_F$ and $\Xc_E$ are
Hopf sualgebras with respect to a coproduct $\Delta^{(D)}$ which differs from the coproduct \eqref{stcop}.
It is sufficient to describe this coalgebraic properties only for the generators of the current Borel subalgebras
and they are
\begin{gather*}
\Delta^{(D)}k^\pm_j(u)=k^\pm_j(u)\otimes k^\pm_j(u),
\\
\Delta^{(D)}F_i(u)=1\otimes F_i(u)+F_i(u)\otimes k^+_i(u)k^+_{i+1}(u)^{-1},
\\
\Delta^{(D)}E_i(u)= E_i(u)\otimes 1+ k^-_i(u)k^-_{i+1}(u)^{-1}\otimes E_i(u),
\end{gather*}
for $\DYGLN$: $i=1,\dots,N-1$, $j=1,\dots,N$;
 for $\DYBn$: $i=0,\dots,n-1$, $j=0,\dots,n$; for~$\DYCn$: $i=0,\dots,n-1$, $j=1,\dots,n$ and
\begin{gather*}
\Delta^{(D)}k^\pm_j(u)=k^\pm_j(u)\otimes k^\pm_j(u),
\\
\Delta^{(D)}F_i(u)=1\otimes F_i(u)+F_i(u)\otimes k^+_i(u)k^+_{i+1+\delta_{i,0}}(u)^{-1},
\\
\Delta^{(D)}E_i(u)= E_i(u)\otimes 1+ k^-_i(u)k^-_{i+1+\delta_{i,0}}(u)^{-1}\otimes E_i(u).
\end{gather*}
 for $\DYDn$: $i=0,\dots,n-1$, $j=1,\dots,n$.
It is clear from these coproduct formulas that current Borel subalgebras $\Xc_F$ and $\Xc_E$ are
Hopf subalgebras in $\DYg$.

Define the intersections
\begin{gather}
\Xc_F^-=\Xc_F\cap \Xc^-,\qquad \Xc_F^+=\Xc_F\cap \Xc^+,\nonumber
\\
\Xc_E^-=\Xc_E\cap \Xc^-,\qquad \Xc_E^+=\Xc_E\cap \Xc^+.
\label{inter}
\end{gather}
According to definition of $\Xc_F$ and $\Xc_E$ we have
\begin{gather*}
\Xc_F^+=\Xc_f^+\cup \Xc_k^+,\qquad
\Xc_F^-=\Xc_f^-,\qquad
\Xc_E^-=\Xc_e^-\cup \Xc_k^-,\qquad
\Xc_E^+=\Xc_e^+.\qquad
\end{gather*}
Each intersections in \eqref{inter} is a subalgebra in the Yangian double $\DYg$ and they are all coideals
with respect to the Drinfeld coproduct $\Delta^{(D)}$
\begin{gather*}
\Delta^{(D)}(\Xc_F^+)=\Xc_F\ot \Xc_F^+,\qquad \Delta^{(D)}(\Xc_F^-)=\Xc_F^-\ot \Xc_F,
\\
\Delta^{(D)}(\Xc_E^+)=\Xc_E\ot \Xc_E^+,\qquad \Delta^{(D)}(\Xc_E^-)=\Xc_E^-\ot \Xc_E.
\end{gather*}

According to the Cartan--Weyl construction of the Yangian double we may impose
a global ordering of the generators in this algebra. There are two different choices
for such an ordering. We denote the ordering relation by the symbol $\prec$ and introduce the cycling
ordering between element of the subalgebras $\Xc^\pm_f$, $\Xc^\pm_e$ and $\Xc_k^\pm$ as follows \cite{EKhP07}
\begin{gather}\label{order}
\cdots \prec \Xc_f^-\prec \Xc_f^+\prec \Xc_{k}^+\prec \Xc_e^+\prec \Xc_e^-\prec \Xc_{k}^-\prec \Xc_f^-\prec \cdots.
\end{gather}
Using this ordering rule we may say that arbitrary elements $\F\in \Xc_F$ and $\E\in \Xc_E$
are ordered if they are presented in the form
\begin{gather*}
\F=\F_-\cdot \F_+,\qquad \E=\E_+\cdot\E_-,
\end{gather*}
where $\F_\pm\in \Xc_F^\pm$ and $\E_\pm\in \Xc_E^\pm$.

According to the general theory one may define projections of any ordered elements from the
subalgebras $\Xc_F$ and $\Xc_E$ onto subalgebras \eqref{inter} using the formulas
\begin{gather}
\Pfp(\F_-\cdot\F_+)=\varepsilon(\F_-)\F_+,\qquad
\Pfm(\F_-\cdot \F_+)=\F_-\varepsilon(\F_+),\qquad
\F_\pm\in \Xc_F^\pm,\nonumber
\\
\Pep(\E_+\cdot\E_-)=\E_+\varepsilon(\E_-),\qquad
\Pem(\E_+\cdot\E_-)=\varepsilon(\E_+)\E_-,\qquad
\E_\pm\in \Xc_E^\pm,
\label{proj1}
\end{gather}
where the counit morphism $\varepsilon\colon \DYg \to \CC$ is defined by the rules
\begin{gather}\label{coun}
\varepsilon(F_i[\ell])=\varepsilon(E_i[\ell])=0,\qquad \varepsilon(k_j[\ell])=0.
\end{gather}

In the next section we will introduce certain elements
in the current Borel subalgebras $\Xc_F$ and $\Xc_E$ given by the product of the
currents at coinciding values of the spectral parameters.
Since the currents $F_i(u)$ and $E_i(u)$ are given by the infinite series \eqref{modcF} and \eqref{modcE}
 we have to assign meaning to such products. This can be done by introducing
certain completions of the current Borel subalgebras $\Xc_F$ and $\Xc_E$
(see \cite{HutLPRS17a} for the description of these completions in case of the Yangian
doubles associated with supersymmetric algebras $\mathfrak{gl}(m|n)$).
Let $\oXc_F$ be the completion of the algebra $\Xc_F$ formed by infinite sums of monomials
that are ordered products $\cF_{i_1}[\ell_1]\cdots \cF_{i_a}[\ell_a]$ with $\ell_1\leq\cdots\leq\ell_a$,
where $\cF_{i_l} [\ell_l]$ is either $F_{i_l}[\ell_l]$ or $k_{i_l}[\ell_l]$. Define analogously~$\oXc_E$ as
the completion of~$\Xc_E$ that are ordered products $\cE_{i_1}[\ell_1]\cdots \cE_{i_b}[\ell_b]$
with $\ell_1\geq\cdots\geq\ell_b$,
where~$\cE_{i_l} [\ell_l]$ is either $E_{i_l}[\ell_l]$ or $k_{i_l}[\ell_l]$.
One can prove that
\begin{enumerate}\itemsep=0pt
\item[(1)] the action of the projections \eqref{proj1} extends to the algebras $\oXc_F$ and $\oXc_E$ respectively;
\item[(2)] for any $\F\in\oXc_F$ with $\Delta^{(D)}(\F)=\F^{(1)}\otimes\F^{(2)}$ we have
\begin{gather}\label{ordF}
\F=\Pfm\big(\F^{(2)}\big)\cdot \Pfp\big(\F^{(1)}\big),
\end{gather}
\item[(3)] for any $\E\in\oXc_E$ with $\Delta^{(D)}(\E)=\E^{(1)}\otimes\E^{(2)}$ we have
\begin{gather}\label{ordE}
\E=\Pep\big(\E^{(1)}\big)\cdot \Pem\big(\E^{(2)}\big).
\end{gather}
\end{enumerate}

Definition of the projections given by the formulas \eqref{proj1} is useful to prove their properties, in particular,
\eqref{ordF} and \eqref{ordE}.
This properties are power tool to calculate the projections of the products of the currents.
But sometimes, one can use more practical way to calculate the projection.
For example, to calculate the projection of the product of the currents $F_i(u)$
one has to replace each current by the combination of the Gauss coordinates
\eqref{DF-iso1} and then use the commutation relations
 to move all ``positive'' Gauss coordinates $\FF^+_{i+1,i}(u)$
 to the right and all ``negative'' Gauss coordinates $\FF^-_{i+1,i}(u)$ to the left. After such reordering
 according to the cyclic ordering \eqref{order}
 the higher Gauss coordinates $\FF^\pm_{j,i}(u)$ with $j>i+1$ will appear
 due to the $RTT$ commutation relations and
 the application of the projection $\Pfp$ amounts to remove
 all the terms containing at least one ``negative'' Gauss coordinate on the left.
 Similarly, the application of the projection $\Pfm$ amounts to remove
 all the terms containing at least one ``positive'' Gauss coordinate on the right.
 The action of the projections $\Pep$ and $\Pem$ is defined analogously according to the
 cyclic ordering \eqref{order} which signifies that Gauss coordinates
 $\EE^+_{i,j}(u)$ should be moved to the left and $\EE^-_{i,j}(u)$ to the right.
 We will use these prescriptions to prove in Appendix~\ref{ApA} Proposition~\ref{cuvsgc} which yields the relations between all Gauss
 coordinates of the $T$-operators $T^\pm(u)$ and the currents generators of
 $\DYg$.

\section{Gauss coordinates and projections of composed currents}\label{GCvscur}

Due to the relations \eqref{oth-ge}, \eqref{sp-ge} and \eqref{o2n-ge} we can introduce
currents $F_i(u)$ and $E_i(u)$ for negative values of the index $i$
\begin{itemize}\itemsep=0pt
\item for $\DYBn$
\begin{gather*}
F_i(u)=-F_{-i-1}(u+c(i+3/2)),\quad E_i(u)=-E_{-i-1}(u+c(i+3/2)),\quad -n\leq i\!\leq -1,
\end{gather*}
\item for $\DYCn$
\begin{gather*}
F_i(u)=-F_{-i}(u+c(i-1)),\qquad E_i(u)=-E_{-i}(u+c(i-1)),\qquad -n+1\leq i\leq -1,
\end{gather*}
\item for $\DYDn$
\begin{gather*}
F_i(u)=-F_{-i}(u+c(i+1)),\qquad E_i(u)=-E_{-i}(u+c(i+1)),\qquad -n+1\leq i\leq -1.
\end{gather*}
\end{itemize}

For the Yangian doubles $\DYGLN$, $\DYBn$ and $\DYCn$ we introduce so called
{\it composed currents}
\begin{gather}\label{toFABC}
\F_{j,i}(u)=F_i(u)\cdot F_{i+1}(u)\cdots F_{j-2}(u)\cdot F_{j-1}(u)\in\oXc_F,
\\
\label{toEABC}
\E_{i,j}(u)=E_{j-1}(u)\cdot E_{j-2}(u)\cdots E_{i+1}(u)\cdot E_{i}(u)\in\oXc_E,
\end{gather}
\looseness=-1
where for $\Xc=\Ac$: $1\leq i<j\leq N$, for $\Xc=\Bc$: $-n\leq i<j\leq n$
and for $\Xc=\Cc$: $-n+1\leq i<j\leq n$.

For the Yangian double $\DYDn$ the picture of the composed currents $\F_{j,i}(u)\in \oDc_F$ and
\mbox{$\E_{i,j}(u)\in \oDc_E$} is more involved and given by the formulas
\begin{gather}\label{comFD}
\F_{j,i}(u)=\begin{cases}
 F_i(u)\cdots F_{j-1}(u),\quad\ -n+1\leq i<j \leq 0\quad\ {\rm and}\quad\ 1\leq i<j \leq n,\\
 -F_i(u)\cdots F_{-2}(u)F_0(u),\quad\ -n+1\leq i \leq -1,\quad\ j=1,\\
 0,\quad\ i=0,\quad\ j=1,\\
 F_0(u)F_2(u)\cdots F_{j-1}(u),\quad\ i=0,\quad\ 2\leq j\leq n,\\
 -F_i(u)\cdots F_{-2}(u)\cdot F_0(u)F_1(u)\cdot F_2(u)\cdots F_{j-1}(u),\quad\
 i\leq -1,\quad\ 2\leq j,
\end{cases}
\end{gather}
and
\begin{gather}\label{comED}
\E_{i,j}(u)=\begin{cases}
E_{j-1}(u)\cdots E_{i}(u),\quad\ -n+1\leq i<j \leq 0\quad\ {\rm and}\quad\ 1\leq i<j \leq n,\\
-E_0(u)E_{-2}(u)\cdots E_{i}(u),\quad\ -n+1\leq i \leq -1,\quad\ j=1,\\
0,\quad\ i=0,\quad\ j=1,\\
E_{j-1}(u)\cdots E_{2}(u)E_0(u),\quad\ i=0,\quad\ 2\leq j\leq n,\\
-E_{j-1}(u)\cdots E_2(u)\cdot E_{1}(u)E_0(u)\cdot E_{-2}(u)\cdots E_{i}(u),\quad\,
 i\leq -1,\quad\, 2\leq j.\!\!\!
\end{cases}
\end{gather}
These formulas looks rather complicated with respect to the formulas \eqref{toFABC} and \eqref{toEABC},
but this is because of restriction \eqref{Drest} and commutativity of the currents $[F_0(u),F_1(v)]=0$
and $[E_0(u),E_1(v)]=0$.
The products $F_i(u)\cdots F_{-2}(u)$,
$F_2(u)\cdots F_{j-1}(u)$ in the second, fourth and fifth lines of \eqref{comFD} disappear for the values of the
indices $i=-1$ and $j=2$. The same is valid for the formula \eqref{comED}.
According to these remarks the composed currents $\F_{2,-1}(u)$ and $\E_{-1,2}(u)$
are equal to $-F_0(u)F_1(u)$ and $-E_0(u)E_1(u)$.

\begin{prop}\label{cuvsgc}
Gauss coordinates of $T$-operators for the Yangian double $\DYg$ associated with Lie algebras
$\mathfrak{g}=\mathfrak{gl}_N$, $\mathfrak{o}_{2n+1}$,
$\mathfrak{sp}_{2n}$ and $\mathfrak{o}_{2n}$ are
related to the composed currents $\F_{j,i}(u)\in\oXc_F$ \eqref{toFABC}, \eqref{comFD} and $\E_{i,j}(u)\in\oXc_E$ \eqref{toEABC}, \eqref{comED} as follows
$(i,j\in I_{\mathfrak{g}}$, $i<j)$
\begin{gather}\label{PpF}
\Pfp\sk{\F_{j,i}(u)}=\FF^+_{j,i}(u),
\\
\label{PmF}
\Pfm\sk{\F_{j,i}(u)}=\tFF^-_{j,i}(u),
\\
\label{PpE}
\Pep\sk{\E_{i,j}(u)}=\EE^+_{i,j}(u),
\\
\label{PmE}
\Pem\sk{\E_{i,j}(u)}=\tEE^-_{i,j}(u).
\end{gather}
\end{prop}

Formulas \eqref{PpF} and \eqref{PmF} were proved in \cite{HutLPRS17a} using definitions of
the composed currents as residues of the product of the simple root currents for the Yangian double
$\mathcal{D}Y(\mathfrak{gl}(m|n))$.
Formulas \eqref{PpE} and \eqref{PmE} can be proved analogously.

In this paper we present another proof of Proposition~\ref{cuvsgc} which uses only
the commutation relations \eqref{rtt-dy} in the Yangian double $\DYg$ and properties
of the projections \eqref{proj1}. The reader can find this proof in Appendix~\ref{ApA}.

Formulas which relate the Gauss coordinates of $T$-operators to the simple root currents
are important in the
application to the quantum integrable models associated with $\mathfrak{g}$-invariant
$R$-matrices. In particular, they are
important in calculation of the action of monodromy entries~$T_{i,j}(z)$ onto off-shell Bethe vectors
in these models. As it was shown in \cite{HLPRS20} this calculation starts from the
action of the right-upper entry~$T_{1,N}(z)$ onto off-shell Bethe vector. Due to the
Gauss decomposition \eqref{GF} or \eqref{GaussB} this action is proportional to the action of the right-upper
Gauss coordinate $\FF^+_{N,1}(u)$ for $\DYGLN$, $\FF^+_{n,-n}(u)$ for $\DYBn$ and
$\FF^+_{n,-n+1}(u)$ for $\DYCn$, $\DYDn$. Results of this paper show that these Gauss
coordinates are
\begin{itemize}\itemsep=0pt
\item for $\DYGLN$
\begin{gather}\label{AA3}
\FF^+_{N,1}(u)=\Pfp\big(F_1(u)F_2(u)\cdots F_{N-2}(u)F_{N-1}(u)\big),
\end{gather}
\item for $\DYBn$
\begin{gather}
\FF^+_{n,-n}(u)= (-1)^n\Pfp\big(F_{n-1}(u-c(n-3/2))\cdots F_{1}(u-c/2)F_0(u+c/2)\nonumber
\\ \phantom{\FF^+_{n,-n}(u)= (-1)^n\Pfp\big(}
{}\times F_0(u)\cdots F_{n-1}(u)\big),
\label{BB3}
\end{gather}
\item for $\DYCn$
\begin{gather}\label{CC3}
\FF^+_{n,-n+1}(u)\!=\! (-1)^{n-1}\Pfp\big(F_{n-1}(u-cn)
\cdots F_{1}(u-2c) F_0(u)F_1(u) \cdots F_{n-1}(u)\big),\!\!\!
\end{gather}
\item for $\DYDn$
\begin{gather}
\FF^+_{n,-n+1}(u)= (-1)^{n-1}\Pfp\big(F_{n-1}(u-c(n-2))\cdots F_{2}(u-c) F_0(u)\nonumber
\\ \phantom{\FF^+_{n,-n+1}(u)= (-1)^{n-1}\Pfp\big(}
\times F_1(u)F_2(u) \cdots F_{n-1}(u)\big).
\label{DDF}
\end{gather}
\end{itemize}

Equality \eqref{AA3} was used in \cite{HLPRS20} to calculate the action of the
monodromy matrix elements~$T_{i,j}(u)$ onto off-shell Bethe vectors for
the integrable models associated with
$\mathfrak{gl}(m|n)$-invariant $R$-mat\-rices.
Analogously, equality \eqref{BB3} was used in the paper \cite{LP20} to describe
the Bethe vectors for the integrable models related to the
$\mathfrak{o}_{2n+1}$-invariant $R$-matrices.
Equalities \eqref{CC3}--\eqref{DDF} can be used to calculate the actions of monodromy entries
onto corresponding off-shell Bethe vectors in the quantum integrable models associated with
$\mathfrak{sp}_{2n}$- and $\mathfrak{o}_{2n}$-invariant $R$-matrices.

\section*{Conclusion}

In this paper we consider a relations between Gauss coordinates of the $T$-operators for the
Yangian doubles of the classical series and the current generators of the same algebras.
To obtain these results we resolve the constraints imposed by the relation
\eqref{ident} onto Gauss
coordinates corresponding to the simple roots of $\mathfrak{g}$ using
results previously obtained in the paper \cite{LPRS19}. This yields the relations
\eqref{oth-ge}, \eqref{sp-ge} and \eqref{o2n-ge} for the simple roots Gauss coordinates
of the $T$-operators for the Yangian doubles associated with the algebras
$\mathfrak{o}_{2n+1}$, $\mathfrak{sp}_{2n}$ and $\mathfrak{o}_{2n}$. These
relations in turn allow to select the algebraically independent set of
generators for each of the Yangian doubles given by the sets
\eqref{choice1}, \eqref{ch1C} and \eqref{ch1D}. Following Ding--Frenkel approach one can
obtain from the $RTT$ commutation relations the ``new'' realizations of the
Yangian doubles $\DYg$ in terms of the total currents corresponding
to the simple roots of the algebra~$\mathfrak{g}$. $RTT$ and ``new'' realizations
of the Yangian doubles are associated with a different choices of the Borel
subalgebras in these algebras which have nontrivial intersections.
Properties of the projections onto these intersections and $RTT$ commutation
relations for the Gauss coordinates finally allow to prove Proposition~\ref{cuvsgc}
and express all Gauss coordinates of the $T$ operators~$T^\pm(u)$ for the Yangian doubles
$\DYBn$, $\DYCn$ and $\DYDn$ through the projections of the corresponding composed currents.
Equality \eqref{ident} means that matrix entries $\hat T^\pm_{i,j}(u)$ are also
expressed through the ``current'' generators. For the Yangian double $\DYAn$
these relations were obtained in the paper \cite{LPRS19} and presented here
by the formulas \eqref{hFF}--\eqref{hEE}.

Relations given by the Proposition~\ref{cuvsgc}
are important for the investigation of the quantum integrable models
associated with $\mathfrak{g}$-invariant $R$-matrices for $\mathfrak{g}=\mathfrak{gl}_N$,
$\mathfrak{o}_{2n+1}$, $\mathfrak{sp}_{2n}$ and $\mathfrak{o}_{2n}$.
Using approach introduced in \cite{EKhP07,KhP-Kyoto}
and further developed in \cite{HutLPRS17a,HLPRS20}
we may express the off-shell Bethe vectors in terms of the current generators
of the corresponding Yangian doubles. These presentations and relations between
$T$-operators matrix entries and the current generators
allow to calculate the action of these entries onto off-shell Bethe vectors.
The action formulas for the upper triangular entries of $T$-operators allows to obtain the
recurrent relations for the Bethe vectors. The action of the diagonal
entries allows to obtain the Bethe equations,
while the action of the lower triangular entries gives possibilities to calculate the scalar
products of the Bethe vectors and form-factors of the local operators in the corresponding
integrable models. Realization of this program for the integrable models associated with
algebras of the classical series and their quantum deformations will be
presented in our forthcoming publications.

\appendix

\section{Proof of Proposition~\ref{cuvsgc}}\label{ApA}

It was mentioned above that the proof of Proposition~\ref{cuvsgc} can be
obtained by calculating the residues of the products of the simple root currents
as it was done in \cite{HutLPRS17a} for the Yangian double $\mathcal{D}Y(\mathfrak{gl}(m|n))$.
Here we propose an alternative proof which
uses only $RTT$ relations and properties of the projections. In the next two
subsections we will prove equalities \eqref{PpF} and \eqref{PmF}. Formulas
\eqref{PpE} and \eqref{PmE} can be proved analogously.

\subsection[Proof of the equality (\ref{PpF}) for DY(o2n+1) and DY(o2n)]{Proof of the equality (\ref{PpF}) for $\boldsymbol{\DYBn}$ and $\boldsymbol{\DYDn}$}

The proofs for the algebras $\DYAn$, $\DYBn$ and $\DYCn$ are very similar. We consider
first the proof of \eqref{PpF} for $\DYBn$.
Using unitarity property of $R$-matrix \eqref{Runi}
we may rewrite the commutation relations \eqref{rtt-dy} in the form
\begin{gather}\label{dy1}
\bigg(1-\frac{c^2}{(u-v)^2}\bigg) \big( \mathbf{I}\otimes T^+(v) \big) \big( T^-(u)\otimes\mathbf{I} \big)= R(u,v) \big( T^-(u)\otimes\mathbf{I} \big) \big( \mathbf{I}\otimes T^+(v) \big) R(v,u) .
\end{gather}

Let us consider the element $(i,j; k,l)$ in this matrix equation
\begin{gather}
\bigg( 1 - \frac{c^2}{(v-u)^2}\bigg) T^+_{k,l}(v) T^-_{i,j}(u) \nonumber
\\ \qquad
{}= T^-_{i,j}(u) T^+_{k,l}(v) + \frac{c}{u-v} T^-_{k,j}(u) T^+_{i,l}(v) -
\frac{c \delta_{i,k'} }{u-v+c \kappa} T^-_{\ell',j}(u) T^+_{\ell,l}(v) \nonumber
\\ \qquad\phantom{=}
+\frac{c}{v-u} \bigg( T^-_{i,l}(u) T^+_{k,j}(v) + \frac{c}{u-v} T^-_{k,l}(u) T^+_{i,j}(v)
- \frac{c \delta_{i,k'} }{u-v+c \kappa} T^-_{\ell',l}(u) T^+_{\ell,j}(v) \bigg)
\label{ttr2}
\\ \qquad\phantom{=}
 - \frac{c \delta_{j,l'} }{v-u+c \kappa} \bigg( T^-_{i,\ell'}(u) T^+_{k,\ell}(v)
+ \frac{c}{u-v} T^-_{k,\ell'}(u) T^+_{i,\ell}(v) - \frac{c \delta_{i,k'} }{u-v+c \kappa} T^-_{\ell', q'}(u) T^+_{\ell,q}(v)\bigg),\nonumber
\end{gather}
where in the right hand side of \eqref{ttr2} we assume summation over repeating indices $-n\leq\ell,q< n$.
Recall that $\ell'=-\ell$.

In order to prove \eqref{PpF} for the Yangian double $\DYBn$
we consider \eqref{ttr2} for values of the indices $l=i=j-1$ and $-n\leq k<j-1\leq n-1$.
Taking into account that $\delta_{j,l'}=\delta_{j,(j-1)'}=0$ for all $-n\leq j\leq n$ we observe
that last line in \eqref{ttr2} disappears and we have
\begin{gather}
\bigg(1 - \frac{c^2}{(v-u)^2}\bigg) T^+_{k,j-1}(v) T^-_{j-1,j}(u) \nonumber
\\ \qquad
{}= T^-_{j-1,j}(u) T^+_{k,j-1}(v) + \frac{c}{u-v} T^-_{k,j}(u) T^+_{j-1,j-1}(v) -
\frac{c \delta_{{ j-1},k'} }{u-v+c \kappa} T^-_{\ell',j}(u) T^+_{\ell,j-1}(v) \label{ttr3}
\\ \qquad\phantom{=}
{}+\frac{c}{v\!-\!u} \bigg( T^-_{j-1,j-1}(u) T^+_{k,j}(v) +\! \frac{c}{u\!-\!v} T^-_{k,j-1}(u) T^+_{j-1,j}(v)
- \frac{c \delta_{{j-1},k'} }{u\!-\!v\!+\!c \kappa} T^-_{\ell',j-1}(u) T^+_{\ell,j}(v)\! \bigg).\nonumber
\end{gather}
Since formula \eqref{PpF} is an equality in the subalgebra $\Bc^+_f$ let us extract from \eqref{ttr3}
all the terms which may belong to this subalgebra. To do this we have to express
in~\eqref{ttr3} all matrix ent\-ries~$T^\pm_{i,j}(u)$ through Gauss coordinates and order all the
terms with respect to the ordering \eqref{order}. Since we are interested in the terms from the
subalgebra $\Bc^+_f$ we can drop in \eqref{ttr3} all the terms which have upper-triangular
entries $T^-_{i,j}(u)$, $i<j$ to obtain
\begin{gather}\label{ttr4}
\bigg( 1 - \frac{c^2}{(v-u)^2}\bigg) T^+_{k,j-1}(v) T^-_{j-1,j}(u)\Big|_{\Bc^+_f}
\\ \qquad
{}= \frac{c}{v\!-\!u} T^-_{j-1,j-1}(u) T^+_{k,j}(v)
- \frac{c \delta_{{j-1},k'} }{u\!-\!v\!+\!c \kappa} \bigg(T^-_{\ell',j}(u) T^+_{\ell,j-1}(v) + \frac{c}{u\!-\!v} T^-_{\ell',j-1}(u) T^+_{\ell,j}(v)\! \bigg)\bigg|_{\Bc^+_f}.\nonumber
\end{gather}

Let us consider the product $T^+_{k,j-1}(v)T^-_{j-1,j}(u)$ multiplied from the right by the inverse product of
the Cartan currents
$k^+_{j-1}(v) k^-_j(u)$ and restricted to the subalgebra $\Bc^+_f$ after normal ordering according to \eqref{order}
\begin{gather}
T^+_{k,j-1}(v)T^-_{j-1,j}(u)k^+_{j-1}(v)^{-1}k^-_j(u)^{-1}\big|_{\Bc^+_f}\nonumber
\\ \qquad
{}=\big(\FF^+_{j-1,k}(v)k^+_{j-1}(v)+\FF^+_{j,k}(v)k^+_j(v)\EE^+_{j-1,j}(v)+\cdots\big)\nonumber
\\ \qquad \phantom{=}
{}\times \big(\FF^-_{j,j-1}(u)k^-_j(u)+\cdots\big)k^+_{j-1}(v)^{-1}k^-_j(u)^{-1}\big|_{\Bc^+_f}\nonumber
\\ \qquad
{}={\frac{c}{u-v}}\Pfp\big(\FF^+_{j-1,k}(v)\FF^-_{j,j-1}(v)-\FF^+_{j-1,k}(v)
\FF^+_{j,j-1}(v)+\FF^+_{j,k}(v)\big)+\cdots.
\label{cal1B}
\end{gather}
Here $\cdots$ in the second and third lines of \eqref{cal1B} stand for the terms which do not contribute
to the subalgebra $\Bc^+_f$ after brackets expanding and ordering according to \eqref{order}. The dots
in the fourth line
 stand for terms which are regular at $u=v$ after normal ordering. To obtain
last line in \eqref{cal1B} we have used the commutation relations between Gauss coordinates
\begin{gather*}
[\EE^+_{j-1,j}(v),\FF^-_{j,j-1}(u)]={\frac{c}{v-u}}
\big(k^-_{j-1}(u)k^-_{j}(u)^{-1}-k^+_{j-1}(v)k^+_{j}(v)^{-1}\big),
\\
k^+_{j-1}(v)\FF^-_{j,j-1}(u)k^+_{j-1}(v)^{-1}=\frac{u-v+c}{u-v}
\FF^-_{j,j-1}(u)+{\frac{c}{v-u}}\FF^+_{j,j-1}(v),\qquad {j\not=1},
\\
k^+_{0}(v)\FF^-_{1,0}(u)k^+_{0}(v)^{-1}=\frac{(v-u+c)(v-u-c/2)}{(v-u)(v-u+c/2)}\FF^-_{1,0}(u)-
-\frac{c}{u-v} \FF^+_{1,0}(v)
\\ \hphantom{k^+_{0}(v)\FF^-_{1,0}(u)k^+_{0}(v)^{-1}=}
{}-\frac{c}{v-u+c/2} \FF^+_{1,0}(v+c/2).
\end{gather*}
Note that index $j$ in these formulas is in the interval $-n+1\leq j\leq n$.
If we multiply the terms in the right hand side of \eqref{ttr4} by $k^+_{j-1}(v)^{-1} k^-_j(u)^{-1}$ we can
check that normal ordering of all these terms with respect to the ordering \eqref{order} cannot produces
a pole of third order at $u=v$. So multiplying \eqref{ttr4} after normal ordering by $(u-v)^3$ and
setting $u=v$ we obtain an equality
\begin{gather}\label{ite1B}
\FF^+_{j,k}(v)=\Pfp\big(\FF^+_{j-1,k}(v)\big(\FF^+_{j,j-1}(v)-\FF^-_{j,j-1}(v)\big)\big)
=\Pfp\big(\FF^+_{j-1,k}(v)\cdot F_{j-1}(v)\big).
\end{gather}
Let us iterate the equality \eqref{ite1B} once to obtain
\begin{gather}\label{ite10B}
\FF^+_{j,k}(v)=\Pfp\big(\Pfp\big(\FF^+_{j-2,k}(v)\cdot F_{j-2}(v)\big)\cdot F_{j-1}(v)\big).
\end{gather}
One important property of the projections described in \cite{EKhP07} and following from \eqref{ordF} is
that for any element $\F\in \Bc^-_f\cup \Bc^+_f$
\begin{gather}\label{prprB}
\Pfp\sk{\F}=\F+\sum_\ell \Pfm\sk{\F'_\ell}\cdot \Pfp\sk{\F''_\ell},
\end{gather}
 where the sum includes only the terms such that $\F'_\ell\in \Bc^-_f\cup \Bc^+_f$ and $\varepsilon(\F'_\ell)=0$.
Here $\varepsilon$ is the counit map in the Yangian double $\DYBn$ defined by \eqref{coun}. It means
that we can replace the element $\Pfp\sk{\FF^+_{j-2,k}(v)\cdot F_{j-2}(v)}$ in \eqref{ite10B}
by the non-ordered element $\FF^+_{j-2,k}(v)\cdot F_{j-2}(v)\in \Bc^-_f\cup \Bc^+_f$ since
all other terms from \eqref{prprB} will be cancelled by the first projection in the right hand side of \eqref{ite10B}.
This equality will now have the form
\begin{gather*}
\FF^+_{j,k}(v)=\Pfp\big(\FF^+_{j-2,k}(v)\cdot F_{j-2}(v) F_{j-1}(v)\big).
\end{gather*}
Further iterations up to ${\rm F}^+_{k+1,k}(v)$ which can be replaced by $F_{k}(v)$
proves equality \eqref{PpF} for all values $j$ and $k$ such that $-n\leq k<j-1\leq n-1$. When $k=j-1$
equality \eqref{PpF} is valid by definition of the currents \eqref{DF-iso1B}.

To prove \eqref{PpF} in case of the Yangian $\DYDn$ and to obtain the structure of the composed
currents given by the equality \eqref{comFD} we proceed as follows. First we observe that this proof for the
composed currents $\F_{j,i}(u)$ with $-n+1\leq i<j \leq 0$ and $1\leq i<j \leq n$ is analogous
to the proof presented above. This is because the collections of the Gauss coordinates
$\FF^\pm_{j,i}(u)$, $\EE^\pm_{i,j}(u)$, $k^\pm_\ell(u)$ either for $1\leq i<j \leq n$, $1\leq\ell\leq n$
or for $-n+1\leq i<j \leq 0$, $-n+1\leq\ell\leq 0$
form the algebras both isomorphic to $\mathcal{D}Y(\mathfrak{gl}_n)$.

Considering the $RTT$ commutation relation for the Yangian double $\DYDn$ in the form~\eqref{dy1}
we can prove as above that a restriction of the element
\begin{gather*}
T^+_{i,-1}(v)T^-_{-1,1}(u)k^+_{-1}(v)^{-1}k^-_1(u)^{-1}\big|_{\Dc^+_f}=0
\end{gather*}
to the subalgebra $\Dc^+_f\subset \DYDn$ vanishes for $-n+1\leq i\leq -2$. This equality implies that
\begin{gather*}
\FF^+_{1,i}(v)=-\Pfp\big( \FF^+_{-1,i}(v)F_0(v)\big)=-\Pfp\sk{F_i(v)\cdots F_{-2}(v)F_0(v)}
\end{gather*}
proving the second line of \eqref{comFD}.

Repeating the same arguments from the vanishing of the element
\begin{gather*}
T^+_{i,1}(v)T^-_{1,2}(u)k^+_{1}(v)^{-1}k^-_2(u)^{-1}\big|_{\Dc^+_f}=0
\end{gather*}
for $-n+1\leq i\leq -1$ we obtain
\begin{gather*}
\FF^+_{2,i}(v)= \Pfp\big(\FF^+_{1,i}(v)F_1(v)\big) = -\Pfp\big(F_i(v)\cdots F_{-2}(v)F_0(v)F_1(v)\big).
\end{gather*}
From the commutativity of the currents $F_0(u)$ and $F_1(v)$ we have, in particular,
that
\begin{gather*}
\FF^+_{2,-1}(v)=-\FF^+_{2,0}(v)\FF^+_{2,1}(v).
\end{gather*}
Continuing calculations for the vanishing in $\Dc^+_f$ element
\begin{gather*}
T^+_{i,2}(v)T^-_{2,3}(u)k^+_{2}(v)^{-1}k^-_3(u)^{-1}\big|_{\Dc^+_f}=0
\end{gather*}
for $-n+1\leq i\leq 0$ we find that
\begin{gather*}
\FF^+_{3,0}(v)=\Pfp\sk{F_0(v)F_2(v)}\qquad\mbox{and}\qquad
\FF^+_{3,i}(v)=-\Pfp\sk{F_i(v)\cdots F_{-2}(v)F_0(v)F_1(v)F_2(v)}
\end{gather*}
for $-n+1\leq i\leq -1$. Repeating the same calculations we can finally find that
\begin{gather*}
\FF^+_{j,0}(v)=\Pfp\sk{F_0(v)F_2(v)\cdots F_{j-1}(v)}
\end{gather*}
and
\begin{gather*}
\FF^+_{j,i}(v)=-\Pfp\sk{F_i(v)\cdots F_{-2}(v)F_0(v)F_1(v)F_2(v)\cdots F_{j-1}(v)}
\end{gather*}
for $-n+1\leq i\leq -1$ and $3\leq j\leq n$ proving \eqref{PpF} for the algebra $\DYDn$
and the structure of the composed currents \eqref{comFD} in this case.

\subsection[Proof of the equality (\ref{PmF})]{Proof of the equality (\ref{PmF})}

To prove equality \eqref{PmF} for the Yangian double $\DYg$ we proceed in analogous way as
it was done in the previous subsection. The only difference is that because of the
cycling ordering of the subalgebras
$\Xc_e^-\prec \Xc_{k}^-\prec \Xc_f^-$ \eqref{order} instead of the commutation relation \eqref{dy1}
we have to use the commutation relation for the
inverse monodromy $\tilde{T}(u)$ \eqref{inverse}
\begin{gather}\label{dy2}
\bigg(1-\frac{c^2}{(u-v)^2}\bigg) \big( \mathbf{I}\otimes \tilde{T}^+(u) \big)
\big(\tilde{T}^-(v)\otimes\mathbf{I} \big)=
 R(u,v) \big( \tilde{T}^-(v)\otimes\mathbf{I}\big) \big(\mathbf{I}\otimes \tilde{T}^+(u) \big) R(v,u).
\end{gather}
To simplify the presentation we consider here the proof of \eqref{PmF} for the Yangian double
$\DYAn$ only.
Let us take in the matrix equation \eqref{dy2} the matrix element $i+1$, $j$; $i$, $i+1$ to obtain\vspace{-.5ex}
\begin{gather}
\bigg(1-\frac{c^2}{(u-v)^2}\bigg) \tilde{T}^+_{i,i+1}(u)\tilde{T}^-_{i+1,j}(v)= \tilde{T}^-_{i+1,j}(v)\tilde{T}^+_{i,i+1}(u)+\frac{c}{u-v}
 \tilde{T}^-_{i,j}(v)\tilde{T}^+_{i+1,i+1}(u)\nonumber
 \\[-.5ex] \hphantom{\bigg(1-\frac{c^2}{(u-v)^2}}
{}+ \frac{c}{v-u} \tilde{T}^-_{i+1,i+1}(v)\tilde{T}^+_{i,j}(u)-\frac{c^2}{(u-v)^2}
 \tilde{T}^-_{i,i+1}(v)\tilde{T}^+_{i+1,j}(u).
\label{al2}
\end{gather}
Now we substitute in \eqref{al2} the Gauss decomposition of the inverse monodromy matrix
\begin{gather*}
\tilde{T}^\pm_{i,j}(u)=\sum_{\ell=1}^{{\rm min}(i,j)}\tEE^\pm_{\ell,i}(u)k^\pm_\ell(u)^{-1}\tFF^\pm_{j,\ell}(u)
\end{gather*}
and order all the terms with respect to the ordering \eqref{order}.

Since equality \eqref{PmF} is an equality in subalgebra $\Ac^-_f$ we
multiply the relation \eqref{al2} from the left by the product $k^+_i(u)k^-_{i+1}(v)$
 and restrict resulting equality to this subalgebra.
Elements from subalgebra $\Ac^-_f$ can appear only in the left hand side of this equality and
\begin{gather}
k^+_i(u)k^-_{i+1}(v) \tilde{T}^+_{i,i+1}(u)\tilde{T}^-_{i+1,j}(v) \big|_{\Ac^-_f}\nonumber
\\ \qquad
{} = k^+_i(u)k^-_{i+1}(v) \big(k^+_i(u)^{-1} \tFF^+_{i+1,i}(u)+\cdots\big)\nonumber
\\ \qquad \phantom{=}
{}\times \big(k^-_{i+1}(v)^{-1} \tFF^-_{j,i+1}(v)+\tEE^-_{i,i+1}(v)k^-_i(v)^{-1}\tFF^-_{j,i}(v)+\cdots\big)\big|_{\Ac^-_f}\nonumber
\\ \qquad
{}=\Pfm\big(f(v,u)\tFF^+_{i+1,i}(u)\tFF^-_{j,i+1}(v)-g(v,u)
\big(\tFF^-_{i+1,i}(v)\tFF^-_{j,i+1}(v)-\tFF^-_{j,i}(v)\big)\big)=0.
\label{cal1}
\end{gather}
Last equality in \eqref{cal1} can be rewritten as
\begin{gather}\label{ite2}
\tFF^-_{j,i}(v)=-\Pfm\big(\big(\tFF^+_{i+1,i}(v)-\tFF^-_{i+1,i}(v)\big) \tFF^-_{j,i+1}(v)\big)=\Pfm\big(F_{i}(v)\cdot \tFF^-_{j,i+1}(v)\big),
\end{gather}
where we used the relation $\tFF^\pm_{i+1,i}(u)=-\FF^\pm_{i+1,i}(u)$.
Iterating \eqref{ite2} and using another property of the projections that
\begin{gather*}
\Pfm\big(\tilde\F)=\tilde\F+\sum_\ell \Pfm\big(\tilde\F'_\ell\big)\cdot \Pfp\big(\tilde\F''_\ell\big)\qquad\mbox{such that }
\varepsilon\big(\tilde\F''_\ell\big)=0\quad \forall\,\ell,
\end{gather*}
we prove relation \eqref{PmF} for the Yangian double $\DYGLN$.

\section[Property of the automorphism (\ref{mapp})]{Property of the automorphism (\ref{mapp})}\label{ApB}

Formulas \eqref{hFF}--\eqref{hEE} describe the automorphism of the Yangian double $\DYGLN$.
We have following
\begin{Lemma}\label{dhat}
Double application of the automorphism \eqref{hFF}--\eqref{hEE} yields the relations
\begin{gather} \label{Gsh}
\hat{\hat\FF}^\pm_{j,i}(u)=\FF^\pm_{j,i}(u-Nc),\qquad \hat{\hat\EE}^\pm_{i,j}(u)=\EE^\pm_{i,j}(u-Nc),\qquad
 \hat{\hat k}^\pm_{i}(u)=k^\pm_{i}(u-Nc),
\end{gather}
which implies that
\begin{gather*}
\hat{\hat T}^\pm_{i,j}(u)=T^\pm_{i,j}(u-Nc).
\end{gather*}
\end{Lemma}

Since Gauss coordinates of $\hTT^\pm_{i,j}(u)$
satisfy the same $RTT$ relations \eqref{rrt2} as $T^\pm_{i,j}(u)$ do and Gauss
coordinates for both monodromy matrices are given by the same formula \eqref{GF},
the Ding--Frenkel isomorphism yields the same commutation relations
\eqref{tkiF}--\eqref{tEF} for the currents $\hF_i(u)$, $\hE_i(u)$ and $\hk^\pm_j(u)$ given by \eqref{map-cur}.
We can repeat all calculations as we did to prove Proposition~\ref{cuvsgc}
to obtain
\begin{gather*}
\hFF^+_{j,i}(u)=\Pfp\big(\hat{\F}_{j,i}(u)\big),\qquad
\hEE^+_{i,j}(u)=\Pep\big(\hat{\E}_{i,j}(u)\big),\qquad 1\leq i<j\leq N,
\end{gather*}
where
\begin{gather*}
\hat{\F}_{j,i}(u)=\hat F_{i}(u)\cdot \hat F_{i+1}(u)\cdots \hat F_{j-2}(u)\cdot \hat F_{j-1}(u)\in\oAc_F,
\\
\hat{\E}_{i,j}(u)=\hat E_{j-1}(u)\cdot \hat E_{j-2}(u)\cdots \hat E_{i+1}(u)\cdot \hat E_{i}(u)\in\oAc_E.
\end{gather*}

Double application of the map \eqref{map-cur} yields
\begin{gather*}
\hat{\hat{F}}_i(u)=-\hat{F}_{N-i}(u-(N-i)c)=F_i(u-Nc),
\\
\hat{\hat{E}}_i(u)=-\hat{E}_{N-i}(u-(N-i)c)=E_i(u-Nc)
\end{gather*}
and
\begin{gather*}
\hat{\hat{\F}}_{j,i}(u)=\F_{j,i}(u-Nc),\qquad \hat{\hat{\E}}_{j,i}(u)=\E_{j,i}(u-Nc).
\end{gather*}
Now first two equalities in \eqref{Gsh} may be easily proved
\begin{gather*}
\hat{\hat{\FF}}^+_{j,i}(u)=\Pfp\big(\hat{\hat{\F}}_{j,i}(u)\big)=\Pfp\sk{\F_{j,i}(u-Nc)}=
\FF^+_{j,i}(u-Nc),
\\
\hat{\hat{\EE}}^+_{i,j}(u)=\Pep\big(\hat{\hat{\E}}_{i,j}(u)\big)=\Pep\sk{\E_{i,j}(u-Nc)}=
\EE^+_{i,j}(u-Nc).
\end{gather*}
The third relation in \eqref{Gsh} follows from the commutation relation \eqref{tEF}. \qed

Note that double application of the map \eqref{mapp} to the $T$-operators for the
Yangian doubles $\DYBn$, $\DYCn$ and $\DYDn$ is much simple due to the relation \eqref{ident}.
Taking into account the values of the parameter $\kappa$ \eqref{kappa} for these algebras we obtain
\begin{itemize}\itemsep=0pt
\item for $\mathfrak{o}_{N}$
\begin{gather*}
\hat{\hat T}^\pm_{i,j}(u)=T^\pm_{i,j}(u-(N-2)c),
\end{gather*}
\item for $\mathfrak{sp}_{2n}$
\begin{gather*}
\hat{\hat T}^\pm_{i,j}(u)=T^\pm_{i,j}(u-(2n+2)c).
\end{gather*}
\end{itemize}

Taking results of Lemma~\ref{dhat} into account we can extend
the statement of Proposition~\ref{cuvsgc} and formulate
\begin{prop}
There are relations between Gauss coordinates and the currents for the Yangian double $\DYGLN$
 \begin{gather}
 \Pfp(\F_{j,i}(u)) = \FF^+_{j,i}(u), \nonumber
 \\
 \Pfp(\hat{\F}_{j,i}(u)) = \hFF^+_{j,i}(u)= \tFF^+_{i',j'}(u - (N+1-j)c) ,\nonumber
 \\
 \Pfm(\F_{j,i}(u)) = \tFF^-_{j,i}(u) = \hFF^-_{i',j'}(u + i c),\nonumber
 \\
 \Pfm(\hat{\F}_{j,i}(u)) = \hat{\hat{{\rm F}}}^-_{i',j'}(u+ic) = \FF^-_{i',j'}(u - (N-i) c),
 \label{Frel}
 \end{gather}
 and
 \begin{gather}
 \Pep(\E_{i,j}(u))= \EE^+_{i,j}(u), \nonumber
 \\
 \Pep(\hat{\E}_{i,j}(u)) =\hEE^+_{i,j}(u)= \tEE^+_{j',i'}(u - (N+1-j)c) ,\nonumber
 \\
 \Pem(\E_{i,j}(u)) = \tEE^-_{i,j}(u) = \hEE^-_{j',i'}(u + i c),\nonumber
 \\
 \Pem(\hat{\E}_{i,j}(u)) = \hat{\hat{{\rm E}}}^-_{j',i'}(u + i c) = \EE^-_{j',i'}(u - (N-i) c).
\label{Erel}
 \end{gather}
\end{prop}

Indeed, first line and first equality of the second line in \eqref{Frel} was proved in Proposition~\ref{cuvsgc}.
The second equality in the second line of~\eqref{Frel} is definition of the mapping~\eqref{hFF}.
The third line in \eqref{hFF} was proved in~\eqref{PmF}. The equalities
in the fourth line is just application of the mapping~$\hat{\ }$
to the third line in \eqref{Frel}.
Formulas \eqref{Erel}
can be proved analogously.

\subsection*{Acknowledgments}
The work was performed at the Steklov Mathematical Institute of Russian Academy of Sciences, Moscow.
This work is supported by the Russian Science Foundation under grant 19-11-00062.
Authors thank E.~Ragoucy and N.A.~Slavnov for fruitful discussions.
The authors are grateful to the anonymous referees for essential comments that allowed us to improve this paper.

\pdfbookmark[1]{References}{ref}
\LastPageEnding

\end{document}